\documentclass[%
reprint,
superscriptaddress,
amsmath,amssymb,
aps,
prstab,
]{revtex4-2}
\usepackage{xcolor}
\usepackage{graphicx}% Include figure files
\usepackage{dcolumn}% Align table columns on decimal point
\usepackage{bm}% bold math
\usepackage{booktabs}% for table tabs
\usepackage{multirow}% for multi-row in tables
\usepackage{amssymb}
\usepackage{float}

\begin{document}

\preprint{APS/Cold work Nb}
\title{Role of microstructure on flux expulsion of superconducting radio frequency cavities}

 \author{B. D. Khanal}
 \affiliation{Department of Physics, Old Dominion University, Norfolk, VA 23529, USA}
    \author{S. Balachandran}
  \affiliation{Thomas Jefferson National Accelerator Facility, Newport News, VA 23606, USA}
\author{S. Chetri}
    \affiliation{Applied Superconductivity Center, NHMFL-FSU, Tallahassee, FL 32310, USA}
\author{M. Barron}
\affiliation{Indiana University-Purdue University of Indianapolis, Indianapolis, IN 46202, USA}
\author{R. Mullinix}
\affiliation{Applied Superconductivity Center, NHMFL-FSU, Tallahassee, FL 32310, USA}
\author{A. Williams}
\affiliation{Applied Superconductivity Center, NHMFL-FSU, Tallahassee, FL 32310, USA}
\author{P. Xu}
\affiliation{Brookhaven National Laboratory, Upton, NY 11973, USA }
\author{ A. Ingrole}
\affiliation{Magnet Science and Technology, NHMFL-FSU, Tallahassee, FL 32309, USA}
\author{P. J. Lee}
\affiliation{Applied Superconductivity Center, NHMFL-FSU, Tallahassee, FL 32310, USA}
\author{G. Ciovati}
  \affiliation{Department of Physics, Old Dominion University, Norfolk, VA 23529, USA}
   \affiliation{Thomas Jefferson National Accelerator Facility, Newport News, VA 23606, USA}
   \author{P. Dhakal}
 \email{dhakal@jlab.org}
  \affiliation{Department of Physics, Old Dominion University, Norfolk, VA 23529, USA}
  \affiliation{Thomas Jefferson National Accelerator Facility, Newport News, VA 23606, USA}

% Authors' institution and/or address\\ This line break forced with \textbackslash\textbackslash

\date{\today}% It is always \today, today,

\begin{abstract}
The trapped residual magnetic flux during the cool-down due to the incomplete Meissner state is a significant source of radio frequency losses in superconducting radio frequency (SRF) cavities. Here, we show a clear correlation between the niobium microstructure in elliptical cavity geometry and flux expulsion behavior. In particular, a traditionally fabricated Nb cavity half cell from an annealed poly-crystalline Nb sheet after an 800 $^\circ$C heat treatment leads to a bi-modal microstructure that ties in with flux trapping and inefficient flux expulsion. This non-uniform microstructure is related to varying strain profiles along the cavity shape. A novel approach to prevent this non uniform microstructure is presented by fabricating a 1.3 GHz single cell Nb cavity with a cold-worked sheet and subsequent heat treatment leading to better flux expulsion after 800 $^\circ$C/3 h. Microstructural evolution by electron backscattered diffraction-orientation imaging microscopy on cavity cutouts, and flux pinning behavior by dc-magnetization on coupon samples confirms a reduction in flux pinning centers with increased heat treatment temperature. The heat treatment temperature dependent mechanical properties and thermal conductivity are reported. The significant impact of cold-work in this study demonstrates clear evidence for the importance of microstructure required for high-performance superconducting cavities with reduced losses caused by magnetic flux trapping.  
\end{abstract}

\maketitle

%\tableofcontents

\section{\label{intro}Introduction}
Superconducting radiofrequency (SRF) cavities are the building blocks of modern particle accelerators for fundamental scientific discoveries, security and defense, energy and medical isotope production, materials research, and quantum information systems \cite{padamsee201750, 10.1063/5.0155213}. In recent years, material research has focused on increasing the $Q_0$ in the medium accelerating gradient range ($\sim$20 MV/m) by lowering the BCS resistance of SRF cavity with "dirty" surface layers by impurity diffusion through the Nb surface (titanium, nitrogen, oxygen)\cite{dhakal13,anna13,dhakal14,dhakal20review, posenPRA, ito, lechner2021rf}. The success of this fundamental R\&D discovery is reflected in the adoption of nitrogen treatment in the LCLS-II protocol for producing hundreds of multi-cell cavities commissioned for LCLS cryomodules \cite{gonnella15}. The quest for the highest $Q_0$ and cryogenic efficiency has focused attention on minimizing the overall surface resistance by surface engineering.  

Traditionally, the starting raw material for forming SRF Nb cavities has been annealed Nb sheet with ﬁne grain sizes $\sim$50 µm (ASTM 4-8). However, cavities have also been fabricated from Nb sheets with medium grain sizes of few mm sizes and large grain with centimeter-scale grain sizes.  Fine-grain sheets have better formability characteristics than larger-grain Nb sheets and are preferred for large production batches. Large grain sheets have performance and cost benefits provided repeatable process design can be achieved \cite{kneisel15, Myneni_2023}. Once fabricated, the cavities receive recipe-based surface treatments to create the dirty superconducting layer on the interior surface of the cavity, which improves the quality factor provided that no magnetic flux is trapped. Experimental evidence, and theoretical studies  indicate that the increase in surface resistance due to trapped magnetic flux due to insufficient flux expulsion during the cavity cool-down through the transition temperature, severely degrades the performance  \cite{vogt13,romanenko14,gonnella16,martinello15,vogt15,kubo16}.
Early studies on poorly flux-expelled cavities show increasing the heat treatment temperature beyond  the traditional 800$^\circ$C/3 h can improve flux expulsion \cite{posen19,sung2023evaluation}. With the increase in heat treatment temperature, the grain size is increased while the density of pinning centers (grain boundaries and dislocations) decrease. Coupon studies of SRF-grade Nb suggest that flux trapping and expulsion are related to microstructural dimensions: a)  large grains are better at expelling flux, and b) regions containing fine grains embedded in large grains preferentially trap magnetic flux in fine grain regions \cite{balachandran21}. Material defects in bulk Nb, such as dislocations and segregation of impurities, provide favorable sites for magnetic flux pinning and contribute to additional losses when exposed to the rf field \cite{claire, claireprab, wang2022investigation}. The Nb surface also plays an important role in pinning and rf losses due to vortices \cite{dhakal20flux}. Generally, impurity-diffused cavities are more vulnerable to vortex-induced losses due to the presence of impurities on the surface of the cavity \cite{gonnella16, posen19, martinello16, checchin18}.
Heat treatment, which leads to recrystallization and grain growth, is the main mechanism for reducing the bulk material defects in an SRF Nb cavity. Choosing a suitable cavity heat treatment to maximize the flux expulsion has been complex for cavities: higher heat treatment temperatures result in reduced Nb strength at room temperature \cite{ciovati2015mechanical}, and different heat treatment temperatures were found for cavities made of Nb sheets from different vendors and batches\cite{gonnella2018industrialization, posen2019role}. Recently we reported that fabricating cavities with a cold-worked Nb sheet leads to improved flux expulsion performance after 800 $^\circ$C in cavities, irrespective of the vendor \cite{khanal:napac2022-weze5}. 

In this manuscript, we show that 800 $^\circ$C/3 h can be a sufficient heat treatment condition for full flux expulsion from an SRF Nb vendor, which previously needed a higher heat treatment temperature. This work provides a pathway for low defect SRF Nb cavities starting from a cold-worked Nb sheet and further heat treating to 800 $^\circ$C/3 h to obtain better flux expulsion in an SRF cavity. The lower heat treatment temperature is preferable to maintain the strength of the Nb cavities at room temperature. We reconfirm that the flux expulsion behavior is a bulk microstructure phenomenon, and spatial distribution of uniform microstructure rather than large average grain sizes is an important factor for flux expulsion. This first result directly correlates bulk cavity microstructure with flux expulsion performance through half-cell coupon studies. To better understand the cold-work Nb, physical and mechanical properties were measured to provide an envelope for the applicability of the new starting state of Nb that can be used to fabricate SRF cavity with better flux expulsion. It is to be noted that the duration of heat treatment is limited to 3 hours at all temperatures throughout this study. 

\section{Cavity Fabrications and Sample Preparation}
\subsection{\label{sec:cavfab}Cavity Fabrications}

\begin{figure}[t]
\includegraphics*[width=45mm]{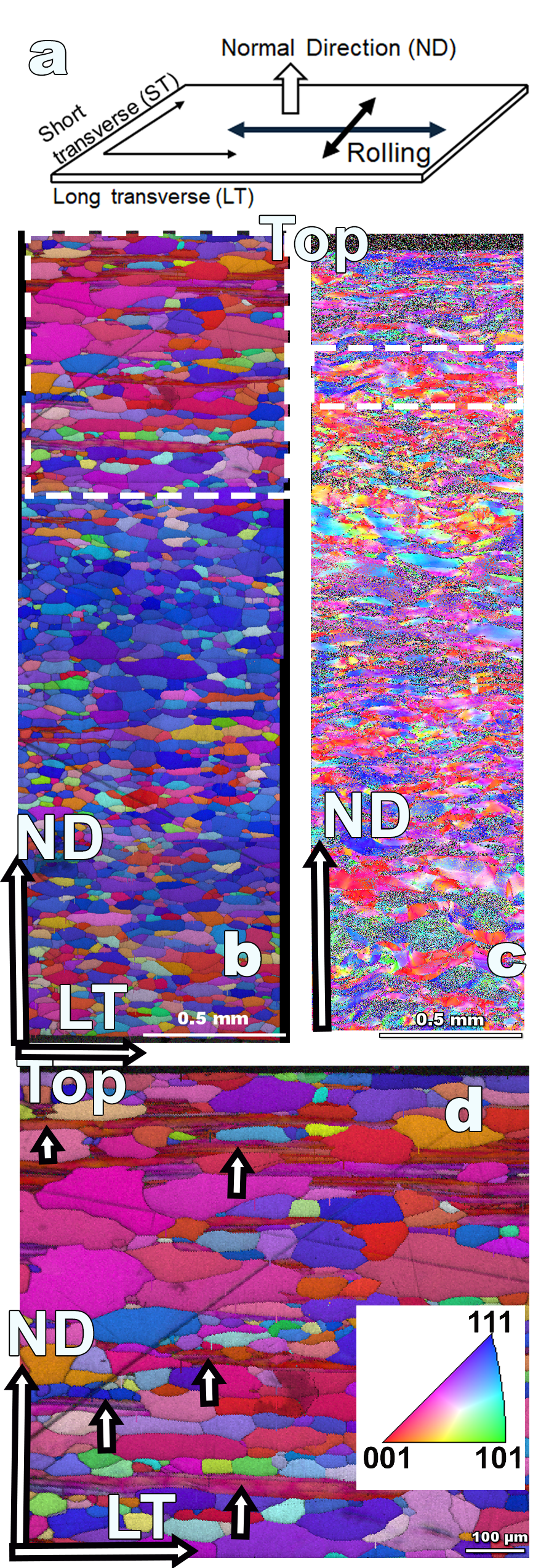}
\caption{\label{fig:ebsd} Initial microstructure of the cross rolled Nb sheets for cavity fabrication a) schematic of rolling direction (b-c) EBSD of SRF-grade and as rolled niobium along ND and (d) near-surface details of SRF grade sheet. }
\end{figure}

\begin{figure}[htb]
\includegraphics*[width=80mm]{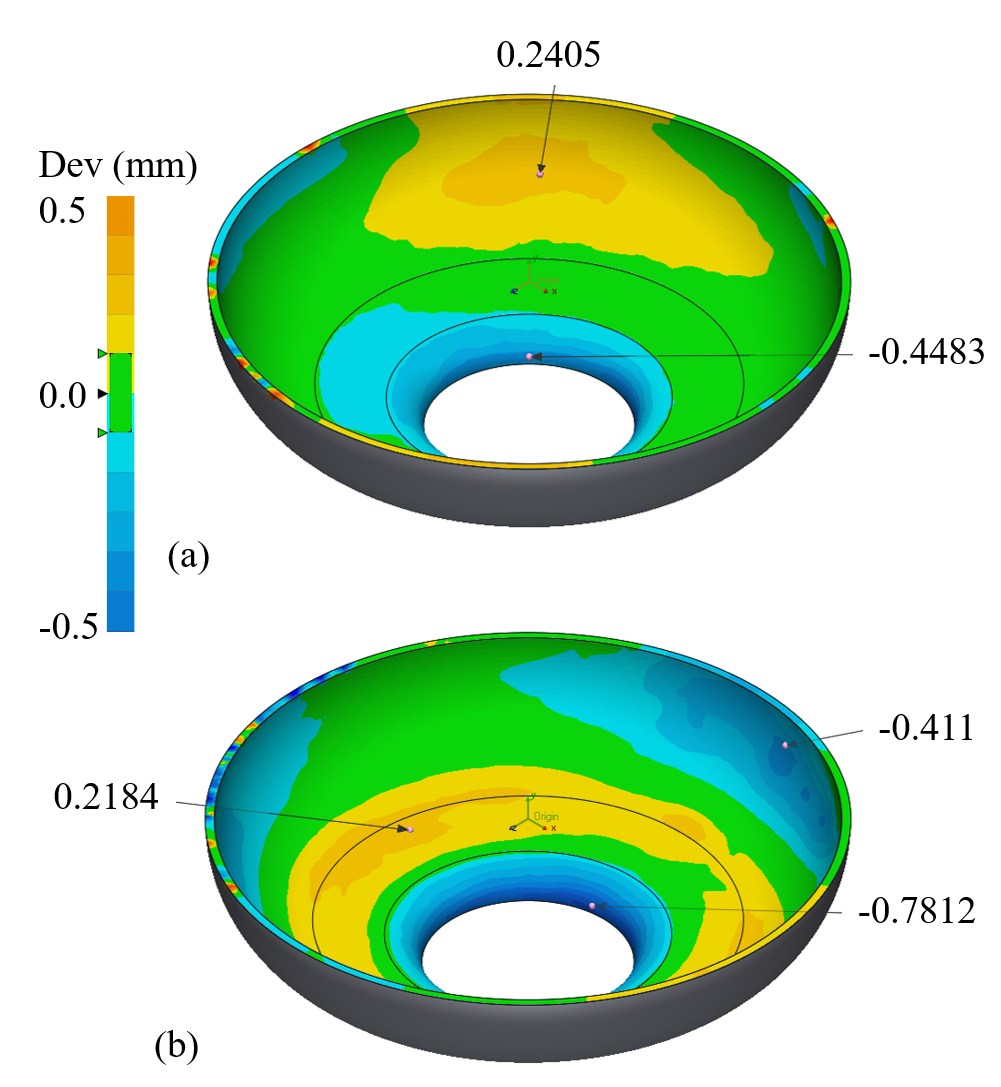}
\caption{\label{fig:3Dscan}3D scan of half-cells and comparison with design for (a) SRF grade and (b) cold-worked Nb.  The region in green corresponds to a deviation within $\pm$0.1 mm.}
\end{figure}

A fine grain Nb sheet of thickness $\sim$ 3 mm was purchased according to the material specification for LCLS-II from the vendor, Tokyo Denkai, Japan. In addition, a special order was made from the vendor to provide sheets in the cold-worked state, with no post-processing by the vendor. The vendor did not provide any material specifications, however, the sheet was made out of the same batch of niobium but without the final annealing step that is normally required to meet traditional SRF Nb sheet specifications. The details about the high purity niobium sheet production steps are presented in Ref. \cite{umezawa2003impurities}. The residual resistivity ratio and the impurity concentrations provided by the vendor are shown in Table \ref{table1}. Electron backscattered diffraction–orientation images (EBSD-OI) acquired for the two sets of Nb sheets are compared in Fig.\ref{fig:ebsd}. For the production of the half-cells, the Nb sheets were machined to discs with outer diameters of 266.7 mm and inner diameters of 47 mm, the same dimensions used for the center cell of 9-cell TESLA-shaped SRF cavity \cite{Aune}. The deep drawing dies available in-house were made from fully anodized aluminum alloy. The half-cells were formed using a male and female die set. The shape deviations of the half-cells were inspected with a 3D scanner and $\sim$85\% of the points were within $\pm$0.2 mm from the ideal shape for SRF-grade and $\sim$71\% for coldworked Nb as shown in Fig. \ref{fig:3Dscan}. A large deviation is seen at the iris region for the cold-worked sheet, which may be corrected by refining the profile of the die for deep drawing. The local shape variation in the iris and equator regions corresponds to 296$\pm$41$\mu$m, and 212$\pm$27$\mu$m, which are in the range of the measurements made with the 3D scanner. 

The fabrication of cavities followed the standard practice of trimming, machining of the iris and equators of the half-cells, and finally electron beam welding of the beam tube (made from low-purity niobium). The regions to be welded were chemically etched by buffered chemical polishing (BCP) to remove $\sim$20 $\mu$m of material after machining. During the equator weld, a failure in the electron beam welding process created holes that were repaired with Nb plugs with an e-beam melt using the Nb material from the available sheet.

\begin{table*}
\caption{\label{table1}
Purity and status of the niobium used for the fabrication of the two single cell cavities used in this study.}
\begin{ruledtabular}
\begin{tabular}{cccccccc}
\textrm{Cavity Name}&
\textrm{Nb Specification, Supplier}&
\textrm{Bulk RRR}&
\textrm{Ta (wt. ppm)}&
\textrm{H (wt. ppm)}&
\textrm{C (wt. ppm)}&
\textrm{O (wt. ppm)}&
\textrm{N (wt. ppm)}\\
\hline
TCA-01 & SRF grade, Tokyo Denkai & 393 & $11$ & $<5$ & $<1$ & $<1$ & $<1$\\
TCNA-01 & Cold work, Tokyo Denkai & 212$^*$ & - & - & - & - & - \\
\end{tabular}

$^*$ RRR is estimated from the thermal conductivity measurements. 
\end{ruledtabular}
 
\end{table*}

After fabrication, the cavities were electropolished (EP) in a horizontal, rotating setup, using a mixture of electronic grade HF:H$_2$SO$_4$ = 1:9 at a constant voltage of $\sim $14 V, a temperature of 15-20 $^\circ$C and a speed of 1~rpm. A total bulk depth of $\sim$150 $\mu$m at cavity equator was removed by EP.
 The first cycle of the heat treatment of the cavity was done at 800~$^\circ$C for 3 hours in a UHV furnace. The cavity was further EP'ed before the rf test. Successive heat treatments of 900 and 1000 $^\circ$C for 3 hours followed by 25~$\mu$m EP were applied before each rf tests.   

\subsection{\label{sec:sampleprep}Coupons Sample Preparation}
Several cylindrical samples of diameter 1~mm and 3~mm in length were machined from the cold-work sheet for the quantification of flux pinning by dc magnetization measurements. One additional half-cell was deep drawn and machined to get strips from the equator to the iris of the cavity. The machined surface of the strips was chemically etched by BCP. The strips and cylinder samples were processed along with the cavities to replicate the processing technique applied to the Nb cavities. Cross-sectional EBSD-OI was performed to evaluate the metallurgical state of the niobium. Additionally, bar samples of 70$\times$3$\times$3 mm$^3$ were machined and processed with high temperature heat treatment to measure the thermal conductivity. Dog-bone shaped samples based on ASTM E8 were machined along, 45$^\circ$ and perpendicular to the cold-rolled direction. Small sample coupons $\sim 10\times6\times3$ ~mm$^3$ were machined and polished to measure the Vickers Hardness (HV) and cross-sectional microstructure measurements. 

\section{\label{ExpRes}Experimental Setup}
\subsection{\label{sec:cooldown}Cavity Test}
Three single-axis cryogenic flux-gate magnetometers (FGM) (Mag-F, Bartington) were mounted on the cavity surface parallel to the cavity axis  $\sim120^{\circ}$ apart in order to measure the residual magnetic flux density at the cavity outer surface during the cooldown process. The magnetic field uniformity within the cavity enclosure is $\sim \pm1$~mG. Six calibrated temperature sensors (Cernox, Lakeshore) were mounted on the cavity: two at the top iris, $\sim180^{\circ}$ apart, two at the bottom iris, $\sim180^{\circ}$ apart, and two at the equator, close to the flux-gate magnetometers. The distance between the temperature sensors at top and bottom iris is $\sim20$~cm. 

The measurement process was in two parts: firstly, several cooldown and warm-up cycles were performed above 10 K in order to explore the flux expulsion ratio while changing the temperature gradient along the cavity axis. The second part of the measurement procedure is as follows: (i) The magnetic field was initially set below $2$~mG using the field compensation coil surrounding the vertical Dewar. (ii) The standard cavity cool-down process was applied, resulting in $<0.1$~K temperature difference between the top and bottom iris. The temperature and magnetic field were recorded until the Dewar was full with liquid He and a uniform temperature of 4.3~K was achieved. This step assumes that all the applied magnetic field is trapped in the SRF cavity walls. (iii) $Q_0(T)$ at low rf field (peak surface rf magnetic field $B_p \sim 20$~mT) from $4.3-1.6$~K was measured using the standard phase-lock technique. (iv) $Q_0$ vs $B_p$ was measured at 2.0 K. (v) The cavity was warmed up above $T_c$ and the residual magnetic field in the Dewar is set to certain values ($\sim$ 20 and $\sim$ 40~mG). Steps (ii) to (iv) were repeated for two different values of magnetic field. 

\subsection{\label{sec:pinning}dc Magnetization}
Isothermal magnetic hysteresis loops were obtained by applying the external dc magnetic field along the axis of the cylinder using a 5T Quantum Design magnetic property measurement system. The applied magnetic field ($B_a=\mu _0H_a$) was then corrected to include the demagnetization and replaced with the effective magnetic field ($B_{eff}= \mu _0H_{eff}$) using the relation, $H_{eff} = H_a-4\pi N M$ (in CGS units) where N is the demagnetization factor which, for a cylindrical sample, is given by \cite{Brandt}:
 \begin{equation}
 N=1-\frac{1}{1+q\dfrac{a}{b}}
 \end{equation}
Here, $q=4/{3\pi}+(2/3\pi) tanh[(1.27(b/a) ln(1+a/b)]$, where $b/a = 3$ is the length to diameter ratio of the cylindrical sample. The pinning force was determined by using the relation ${F_p=B_{eff} \times J_c}$, where $J_c$ is the critical current density obtained from the magnetic hysteresis loop using the Bean model and given by $J_c = 15\Delta M/r$, where $\Delta M$ is the difference in magnetization for $B_{eff}$ during increasing and decreasing field in the hysteresis loop, and $r$ is the radius of the cylinder.
\subsection{\label{sec:thermal}Thermal Conductivity}
Experimental thermal conductivity measurements were performed using an in-house setup and procedures developed for the measurement of thermal conductivity of Nb in the temperature range of 1.8 - 6~K as described in Ref. \cite{gigi}. The apparatus is immersed in liquid helium and lower temperatures were obtained by pumping the liquid helium to sub-atmospheric pressure. A heater was installed at one end of the sample and the other end was exposed to superfluid helium.
In steady state, the thermal conductivity, $\kappa$ of the specimen as a function of temperature ($T$) is given by 
 \begin{equation}
    \kappa = \frac {P d}{A \Delta T}
 \end{equation}
where $P$ is the power supplied to the resistive heater, $d$ is the distance between two temperature sensors installed on the specimen, $\Delta T = T_1-T_2$ is the temperature difference between two temperature sensors and $A$ is the cross-sectional area of the specimen.
\subsection{Microstructural Analysis}
The cross-sectional microstructure of the Nb samples was analyzed using a scanning electron microscope (SEM). Samples were prepared for metallographic analysis by hot-mounting using a conductive bakelite compound, Konductomet\textsuperscript{\textregistered}. The samples were sequentially ground down to obtain flat surfaces with 320, 400, 600, 800, and 1200-grit SiC pads. Diamond polishing with sequential grit sizes of 5~$\mu$m, 3~$\mu$m and 1~$\mu$m, were used for subsequent polishing followed by vibratory polishing using a Vibromet II\textsuperscript{\textregistered} using a 50~nm colloidal silica solution (Mastermet\textsuperscript{\textregistered})with a pH of 11.5. An intermediate light chemical polishing with a BCP 1:1:2, hydrofluoric acid (HF): nitric acid (HNO$_3$): phosphoric acid (H$_3$PO$_4$) was performed; the final step involved re-polishing in the Vibromet\textsuperscript{\textregistered} to achieve damage-free cross-sections. 
Orientation imaging microscopy (OIM) was performed with an EDAX Velocity\textsuperscript{\textregistered} camera capable of greater than 2000 indexed points per second and angular precision below 0.1$^\circ$ in a LaB$_6$ filament, Tescan Vega 3 microscope. Data analysis was performed using EDAX-TSL-OIM software version 8.5.1002.

\subsection{Tensile Testing}
Tensile samples were machined using electrical discharge machining (EDM) from the non-annealed (NA) sheet along the rolling direction (RD), at 45$^\circ$, and 90$^\circ$ to RD. Mechanical behavior, including stress and in-plane strain behavior, was recorded for the Nb sheet to evaluate formability. Only the modulus and strength of the sheets will be presented here. The tensile samples followed the sub-size specimen, ASTM-E8 standard. Annealing corresponding to cavity heat treatments of 800~$^\circ$C, 900~$^\circ$C, and 1000~$^\circ$C were performed on the tensile samples in a UHV furnace. Four tensile samples with and without heat treatment were tested in a MTS Criterion station, with a load cell with 5~kN capacity, under displacement control at a strain rate of 5$\times10^{-5}$~s$^{-1}$. A Class B, 25 mm initial gauge length tensile extensometer with a strain range of 20 \% was used to record the longitudinal strain data. A digital image correlation (DIC) system measured the deformation along the length and thickness. The tensile tests were stopped at criteria when the load dropped by 10 \% of the maximum load reached. This also corresponds to a well-developed neck in the specimen after which instabilities dominate the local deformation. 
The yield strength ($\sigma_{y,0.002}$), referred to as $\sigma_y$ or YS, and ultimate tensile strength (UTS) are reported based on the engineering stress values based on the initial cross-sectional area.

\subsection{Vicker Hardness and Recrystallization}

Samples of nominal dimensions $\sim$ 6$\times$4$\times$3~mm$^3$ were cut from the non-annealed sheet for the heat treatment study. Heat treatments were performed at 300 - 1000~$^\circ$C. The ramp rate used for the study was 5~$^\circ$C/min. The heat treatments between 300 - 700~$^\circ$C were performed by sealing the samples in an evacuated quartz tube with high-purity Ar (Airgas UHP-400). The 800 - 1000~$^\circ$C samples were heat treated in a vacuum furnace used in cavity heat treatments. 
After heat treatments, Vickers microhardness (HV) tests were conducted on the polished cross-section samples using a Starrett Micro Vickers/Knoop Hardness Tester. The hardness results reported here were conducted using an indenter load of 300~g, and a dwell time of 15~seconds, per ASTM E384-05a \cite{ASTM}. 

\subsection{Half-cell Cutout}
During the cavity fabrication, two additional half-cells were deep drawn, one of each from the cold work and SRF grade sheets. The half-cells were cut in strips from equator to the iris to evaluate the cross-sectional microstructure. The strips were chemically etched $\sim$ 50~$\mu$m followed by the UHV heat treatment at 800, 900, and 1000~$^{\circ}$C. The strips were cut into sections that were hot-mounted using a conductive bakelite compound and polished using the sample procedure described above. Electron microscopy and Vickers hardness were measured along the cross-section of the samples.

\section{\label{ExpRes}Cavity Test Results}

\subsection{\label{sec:cooldown}Cool-down and Flux Expulsion}

The ratio of the residual dc magnetic field measured after ($B_{sc}$) and before ($B_n$) the superconducting transition qualitatively explains the effectiveness of the flux expulsion during the transition. A value of $B_{sc}/B_n = 1$ represents complete trapping of magnetic field during cool-down, whereas a flux expulsion ratio of $\sim$1.7 at the equator would result from the ideal superconducting state. Experimentally, $B_{sc}/B_n$ depends on the Nb material and on the temperature gradient along the cavity axis during the cool-down. Values of $B_{sc}/B_n$ close to the theoretical estimate could be achieved with a high temperature gradient ($dT/ds$ $>$ 0.3~K$\cdot$cm$^{-1}$).
\begin{figure}[htb]
\includegraphics*[width=85mm]{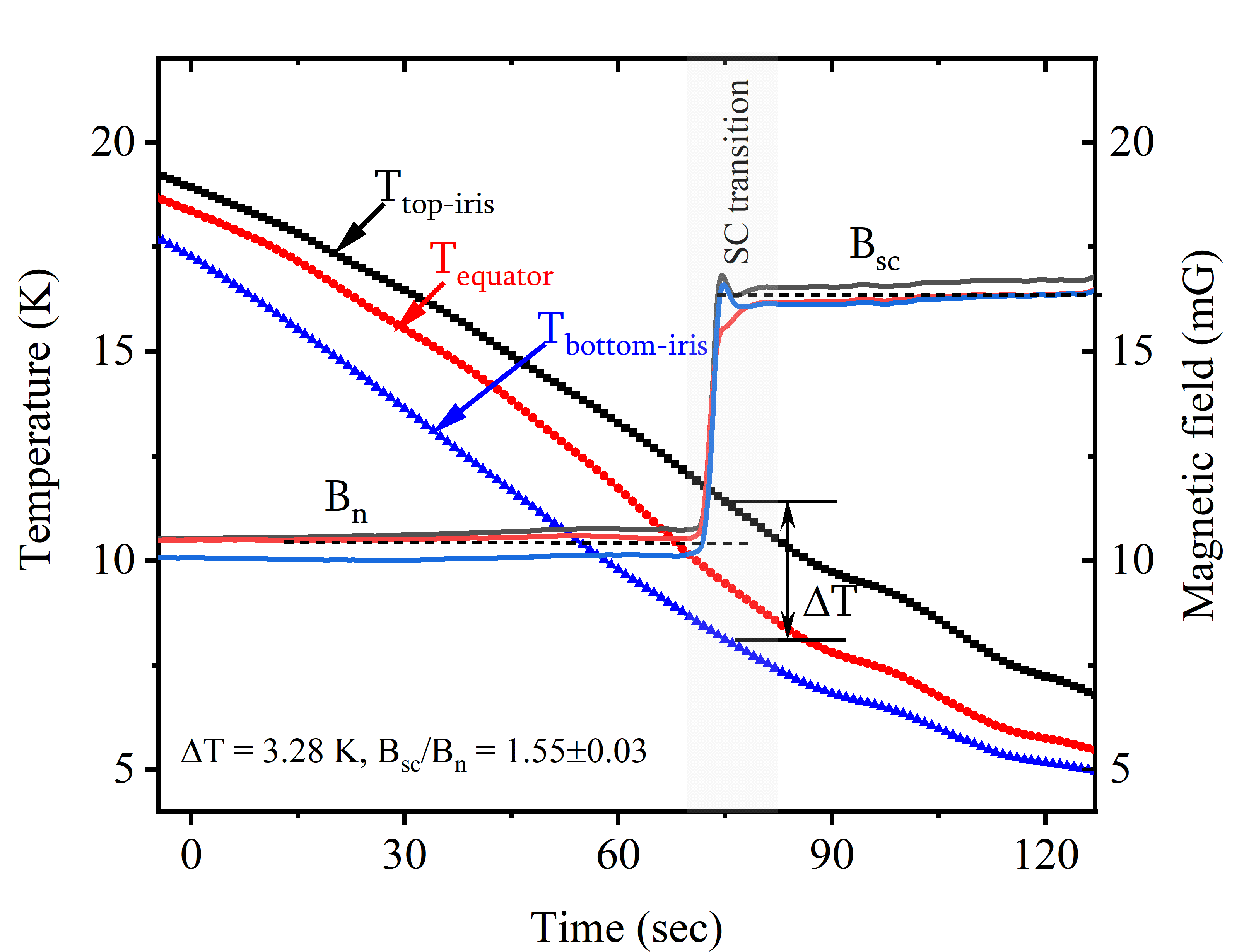}
\caption{\label{fig:transition} Temperature and magnetic field during transition from normal to superconducting state measured during a cool-down cycle of cavity TCA-01.}
\end{figure}
A representative plot of the residual magnetic field at the FGM's locations measured during one cool-down cycle for cavity TCA-01 is shown in Fig.~\ref{fig:transition}. The average value of $B_{sc}/B_n$ for the three FGMs at the equator was $1.55\pm0.03$. The temperature difference between the top and bottom iris when the equator of the cavity reached the superconducting transition temperature $\sim$ 9.25~K is 3.28~K.

\begin{figure}[htb]
\includegraphics*[width=85mm]{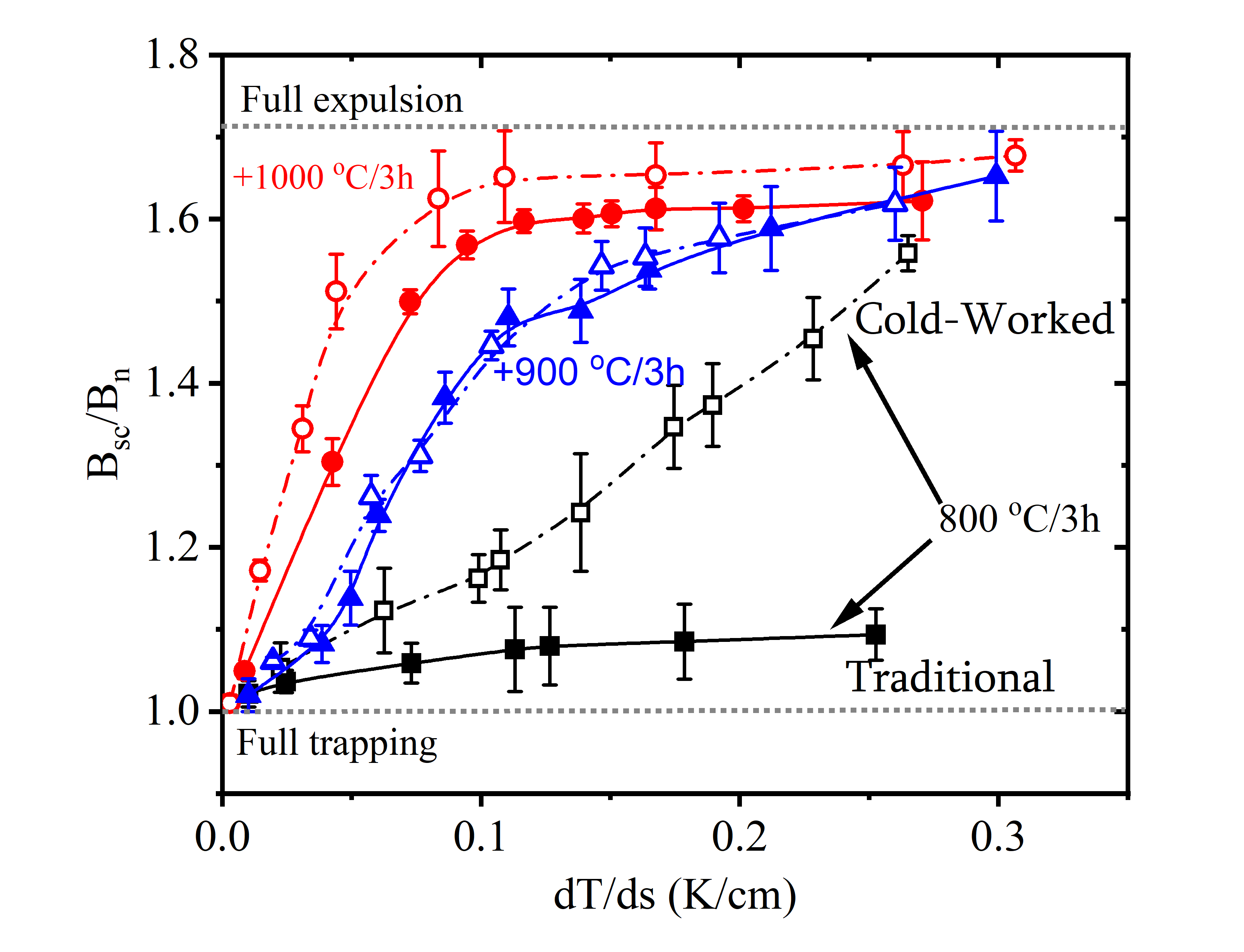}
\caption{\label{fig:expulsion} Average flux expulsion ratio at the equator of SRF cavity as a function of the temperature gradient (iris-to-iris) after each cavity treatments for TCA-01 made from  SRF grade traditional Nb (solid symbols) and TCNA-01 made from cold worked Nb (open symbols). The lines are guide to the eye.}
\end{figure}

Figure \ref{fig:expulsion} shows the flux expulsion ratio $B_{sc}/B_n$ for both cavities with respect to the temperature gradient after each heat treatment. The flux expulsion for cavity made from cold work Nb showed a better expulsion for the cavity heat treatment at 800~$^\circ$C compared to SRF grade Nb. The flux expulsion ratio is similar for both cavities when heat treated with an additional 900~$^\circ$C. However, the cavity made from cold-worked sheet showed better flux expulsion after additional 1000~$^\circ$C heat treatment. The dependence of flux expulsion on temperature difference showed different characteristics depending on the heat treatment temperature. 
 
\subsection{\label{sec:rfdata}RF Measurements}
The average rf surface resistance was obtained from the measurement of $Q_0(T)$ at low rf field ($B_p \sim 20$~mT) for different applied dc magnetic fields, $B_n$, before each cool-down. The data were fitted with the following equation:
\begin{equation}
R_s(T) = R_{BCS}(T,\omega,l,\Delta) + R_{res},
\label{eq1}
\end{equation}
where the BCS surface resistance $R_{BCS}$ was computed numerically from the Mattis-Bardeen (M-B) theory~\cite{MB58} using Halbritter's code~\cite{halbritter70}. The mean free path, $l$, and the ratio $\Delta/k_BT_c$ were regarded as fit parameters, where $\Delta$ is the energy gap at $T=0$~K, and $k_B$ is the Boltzmann constant. We used a coherence length of $\xi _0 = 39$~nm and the London penetration depth, $\lambda_0 = 32$~nm for Nb in the clean limit, $\xi_0\ll l$ at $T=0$. The residual resistance $R_{res}$ was extracted for different trapped magnetic fields. The $T_c$ of the cavity was measured during the cavity warm up. 

\begin{figure}[htb]
\includegraphics*[width=85mm]{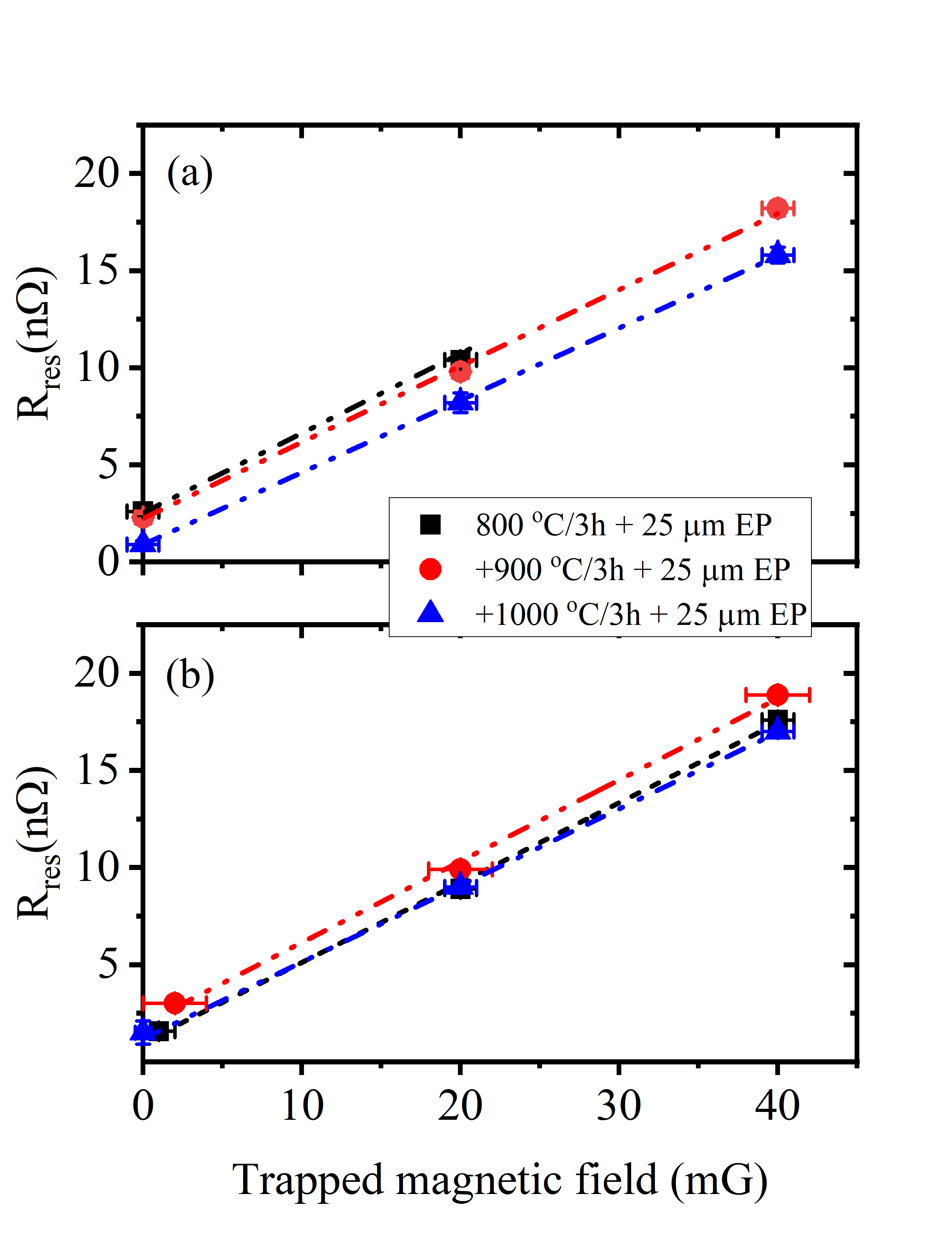}
\caption{\label{fig:RsB} The residual resistance as a function of trapped field after each treatment stage for cavity (a) TCA-01 (traditional sheet) and (b) TCNA-01. }
\end{figure}

Figure \ref{fig:RsB} shows the residual resistance as a function of the trapped dc magnetic field before the cavity transitions from the normal to superconducting state, $B_n$, in full flux trapping condition, $B_{sc}/B_n \sim1$, where $\bigtriangleup T < $~0.1~K. 

\begin{figure}[htb]
\includegraphics*[width=80mm]{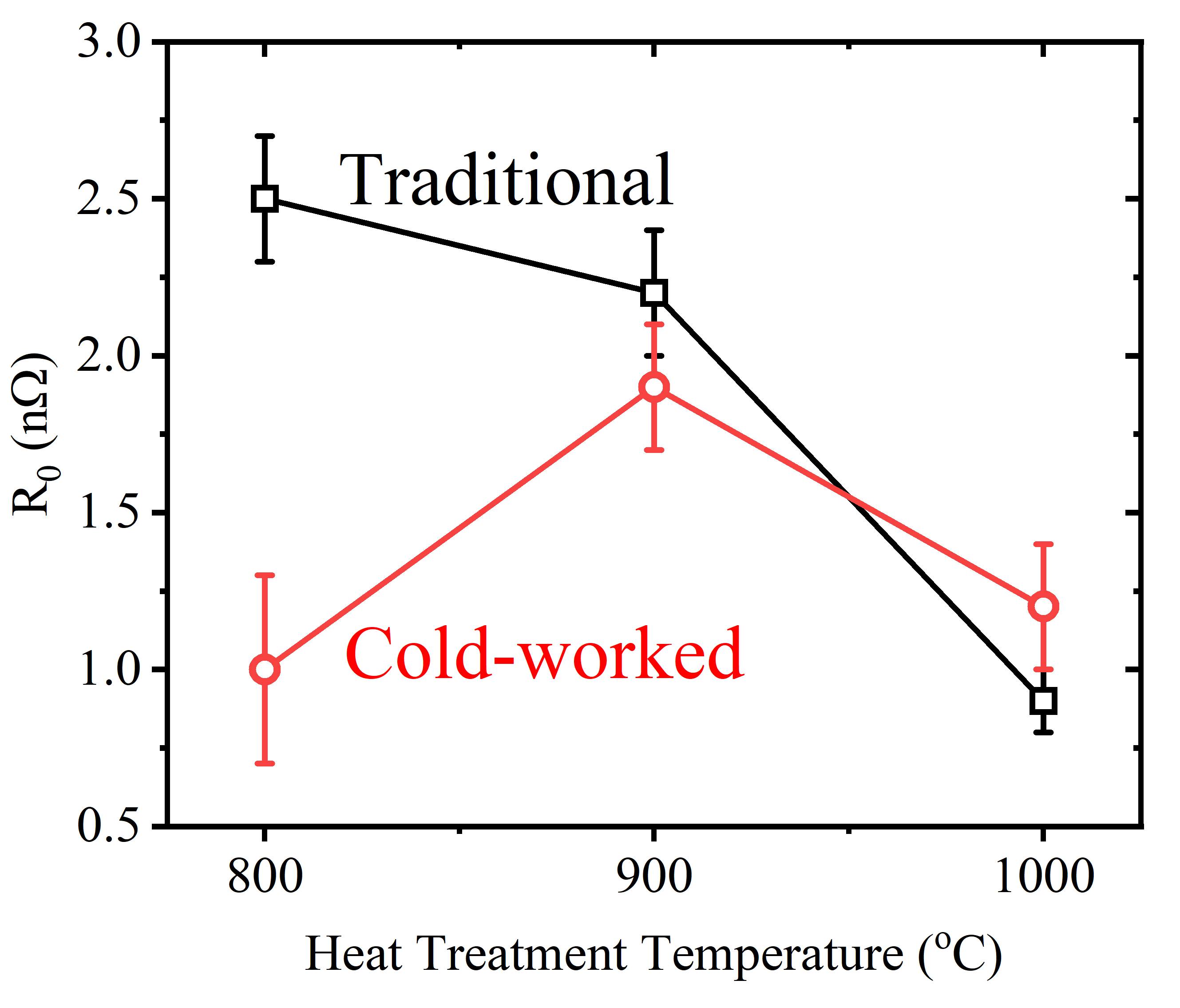}
\caption{\label{fig:RsT} The intrinsic residual resistance due to non-superconducting nano-precipitates, sub-oxide layers at the surface, broadening of the density of states \cite{gurevich17} for cavity TCA-01 and TCNA-01 after each heat treatment followed by $\sim$ 25~$\mu$m EP.}
\end{figure}

The residual resistance due to trapped flux can be written as:
\begin{equation}
R_{res}(B_n) =R_{0}+\eta_t S B_n,
\label{eq2}
\end{equation}
where $R_0$ accounts for the contribution to the residual resistance other than trapped flux, such as non-superconducting nano-precipitates, sub-oxide layers at the surface, broadening of the density of states \cite{gurevich17} is plotted in Fig.\ref{fig:RsT}. $\eta_t$ is the fraction of flux trapping and for $\bigtriangleup T < $~0.1~K, the value of $\eta_t$ $\sim$ 1 and S gives the flux trapping sensitivity. The slope of $R_{res}(B_n)$ for both cavities after heat treatment followed by $\sim$ 25 $\mu$m EP is approximately $\sim$ 0.4~n$\Omega$/mG as shown in Fig. \ref{fig:RsB}.

\begin{figure}[htb]
\includegraphics*[width=85mm]{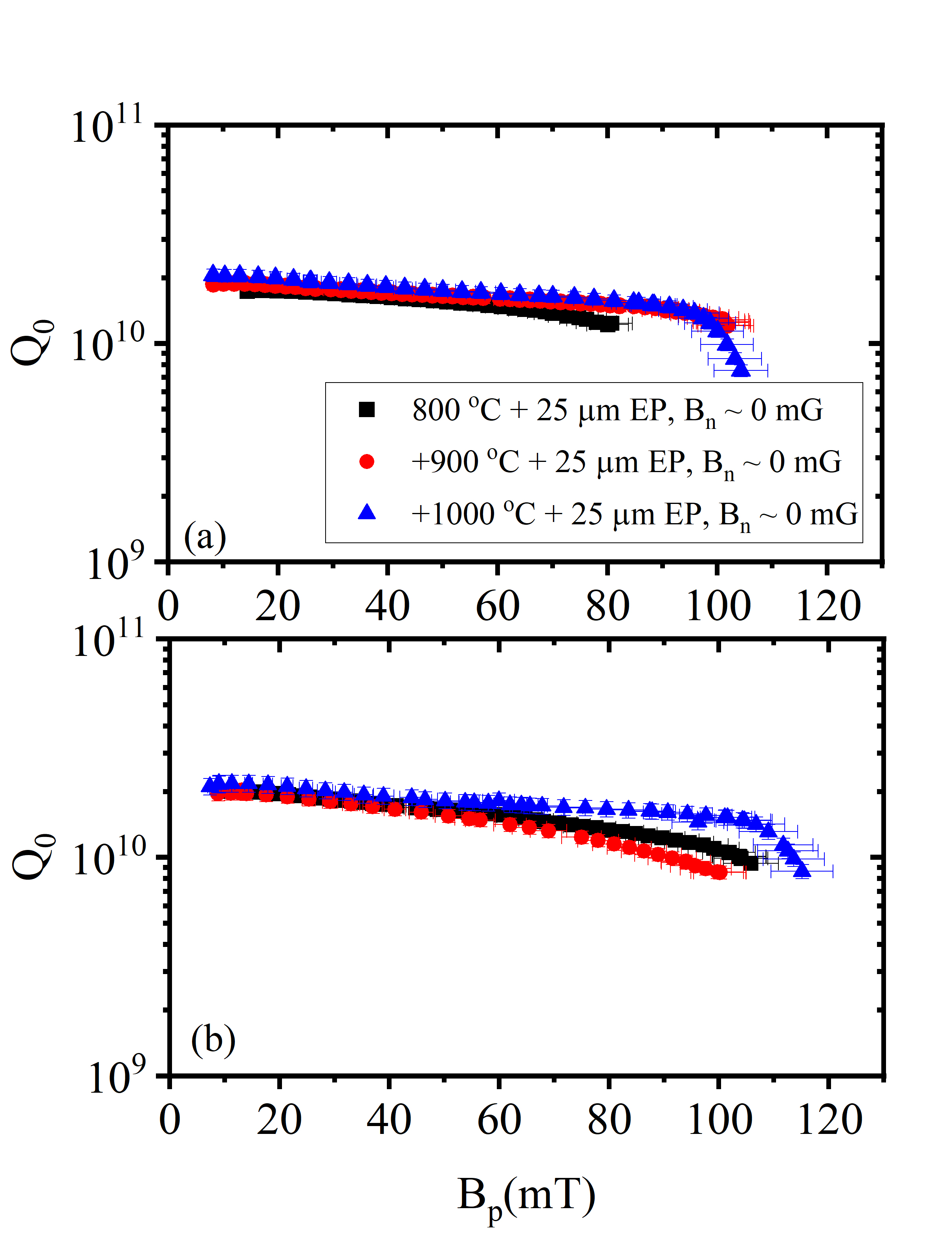}
\caption{\label{fig:QE} $Q_0(B_p)$ measured at 2.0~K after cool-down with $dT/ds$ $>$ 0.2~K$\cdot$cm$^{-1}$ and $B_n < 0$~mG for cavity (a) TCA-01 and (b) TCNA-01. All rf tests were limited by quench.}
\end{figure}

The $Q_0(B_p)$ measured at 2.0~K after cool-down with $dT/ds$ $>$ 0.2~K$\cdot$cm$^{-1}$ and $B_n \sim 0$~mG for each cavity and treatment is shown in Fig.~\ref{fig:QE}. All rf tests were limited by quench and the low quench field may be attributed to the fabrication issue we encountered during cavity welding.

\begin{figure}[htb]
\includegraphics*[width=85mm]{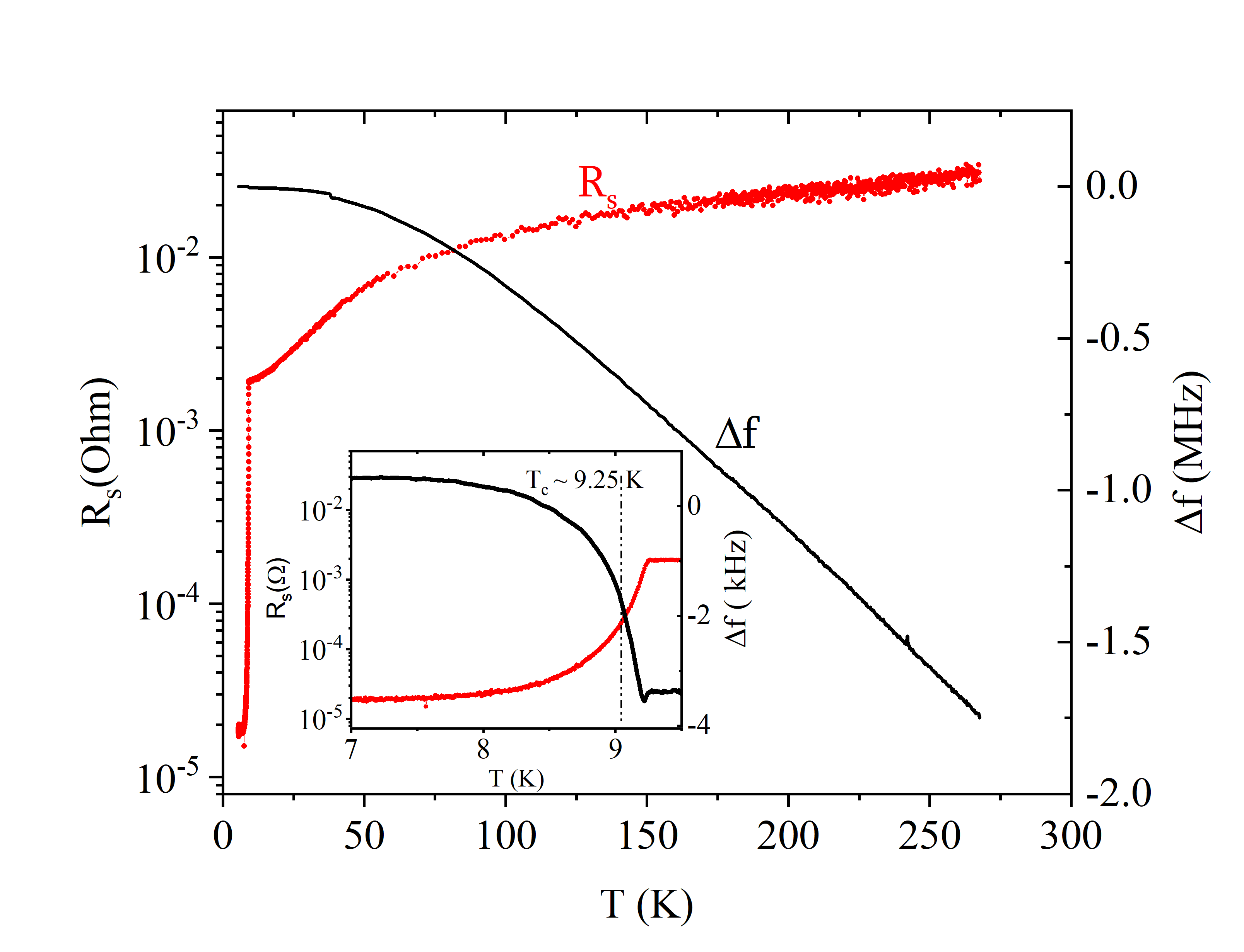}
\caption{\label{fig:RQvsT} The surface resistance and change in resonant frequency as a function of temperature during cavity warm up. The inset shows the superconducting to normal conducting transition at $T_c \sim$ 9.25~K.}
\end{figure}

In order to obtain information about the mean free path near the surface, we measured the resonant frequency and quality factor while warming up the cavities from $\sim$ 5~K to higher than the transition temperature ($>$ 9.3~K) using a vector-network analyzer, from which $R_s(T)$ and the change in resonant frequency was extracted. A representative plot of surface resistance and change in frequency during warm-up is shown in Fig. \ref{fig:RQvsT}. 
The frequency shift can be translated into a change in penetration depth according to 
  \begin{equation}
 \Delta \lambda=\dfrac{G}{\pi \mu_0 f^2}\Delta f
\label{eq3}
\end{equation}
with G being the geometric factor of the cavity, $f$ as the resonant frequency. Using the Casimir-Gorter relation \cite{gorter35}, we can obtained $\lambda_0$, which is the penetration depth at 0~K as:
\begin{equation}
 \Delta \lambda= \lambda(T)-\lambda_0= \dfrac{\lambda_0}{\sqrt{(1-(T/T_c)^4)}} - \lambda_0
\label{eq4}
\end{equation}
In the Pippard limit \cite{pippard53}, $\lambda_0$ is directly related to the mean free path of the quasi-particles as:
\begin{equation}
 \lambda_0= \lambda_L \sqrt{1+\frac{\pi \xi_0}{2l}}.
\label{eq6}
\end{equation}
\begin{figure}[htb]
\includegraphics*[width=85mm]{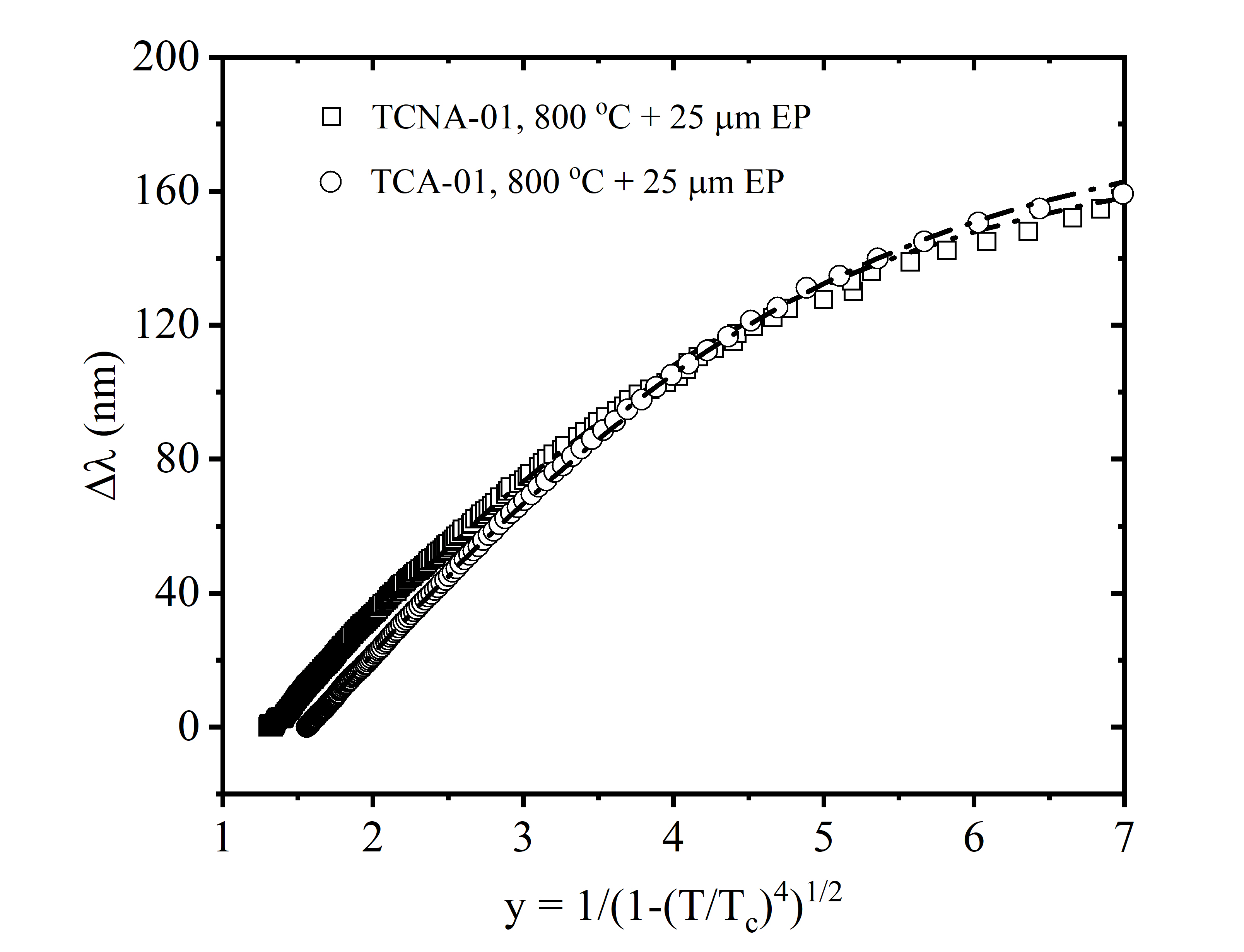}
\caption{\label{fig:lambda} Change of penetration depth as a function of the reduced temperature parameter $y=1/\sqrt{1-(T/T_c)^4}$ measured on cavity TCNA-01 and TCA-01 after 800~$^\circ$C heat treatment followed by $\sim25 \mu$m EP. The dash-dotted lines are fit with M-B theory.}
\end{figure}
The value of the London penetration depth $\lambda_L$ and coherence length of the Cooper pairs $\xi_0$ taken are the same that were used in the $R_s(T)$ fit. Figure \ref{fig:lambda} shows the $\Delta \lambda_L$ vs reduced temperature, y. The data in the superconducting state were fitted using the numerical solution of M-B theory. Table \ref{table2} summarizes all the results of the rf measurements. 
\begin{table*}
\caption{\label{table2}
The flux trapping sensitivity $S$ obtained from fits of $R_{res}(B_n)$ for different treatments and weighted average values of $\Delta /k_BT_c$ obtained from fits of $R_s(T)$ between $1.6 - 4.3$~K for each cavity. The mean free path $l$ was obtained from the fit of $\Delta \lambda_L$ (T) and $T_c$ was measured during the cavity warm up.}
\begin{ruledtabular}
\begin{tabular}{cccccc}
\textrm{Cavity Name}&
\textrm{Treatment}&
\textrm{$S$ (n$\Omega$/mG)}&
\textrm{$\Delta /k_BT_c$}&
\textrm{$l (7.5-9.25$~K) (nm)}&
\textrm{$T_c$ (K)}\\
\hline
 & 800~$^\circ$C + 25~$\mu$m EP & $0.44 \pm 0.01$ & $1.78 \pm 0.02$ & 315 $\pm 2$ & 9.25~ $\pm$ 0.05 \\
TCA-01 & +900~$^\circ$C + 25~$\mu$m EP & $0.43 \pm 0.01$ & $1.80 \pm 0.02$ & 296 $\pm 2$ & 9.24 $\pm$ 0.02 \\
 & +1000~$^\circ$C + 25~$\mu$m EP & $0.40 \pm 0.01$ & $1.80 \pm 0.02$ & 386 $ \pm 9$ & 9.25 $\pm$ 0.03 \\
 \hline
 & 800~$^\circ$C + 25~$\mu$m EP & $0.45 \pm 0.01$ & $1.79 \pm 0.02$ & 375 $\pm 6$ & 9.26 $\pm$ 0.03 \\
TCNA-01 & +900~$^\circ$C + 25~$\mu$m EP & $0.43 \pm 0.01$ & $1.81 \pm 0.02$ & 389 $\pm 6$ & 9.25 $\pm$ 0.01 \\
 & +1000~$^\circ$C + 25~$\mu$m EP & $0.40 \pm 0.01$ & $1.80 \pm 0.02$ & 475 $\pm 4$ & 9.23 $\pm$ 0.03 \\
\end{tabular}
\end{ruledtabular}
\end{table*}

\section{ Analysis of deep-drawn half-cell }
\subsection{\label{sec:half-cell}Hardness profile}
\begin{figure}[htb]
\includegraphics*[width=85mm]{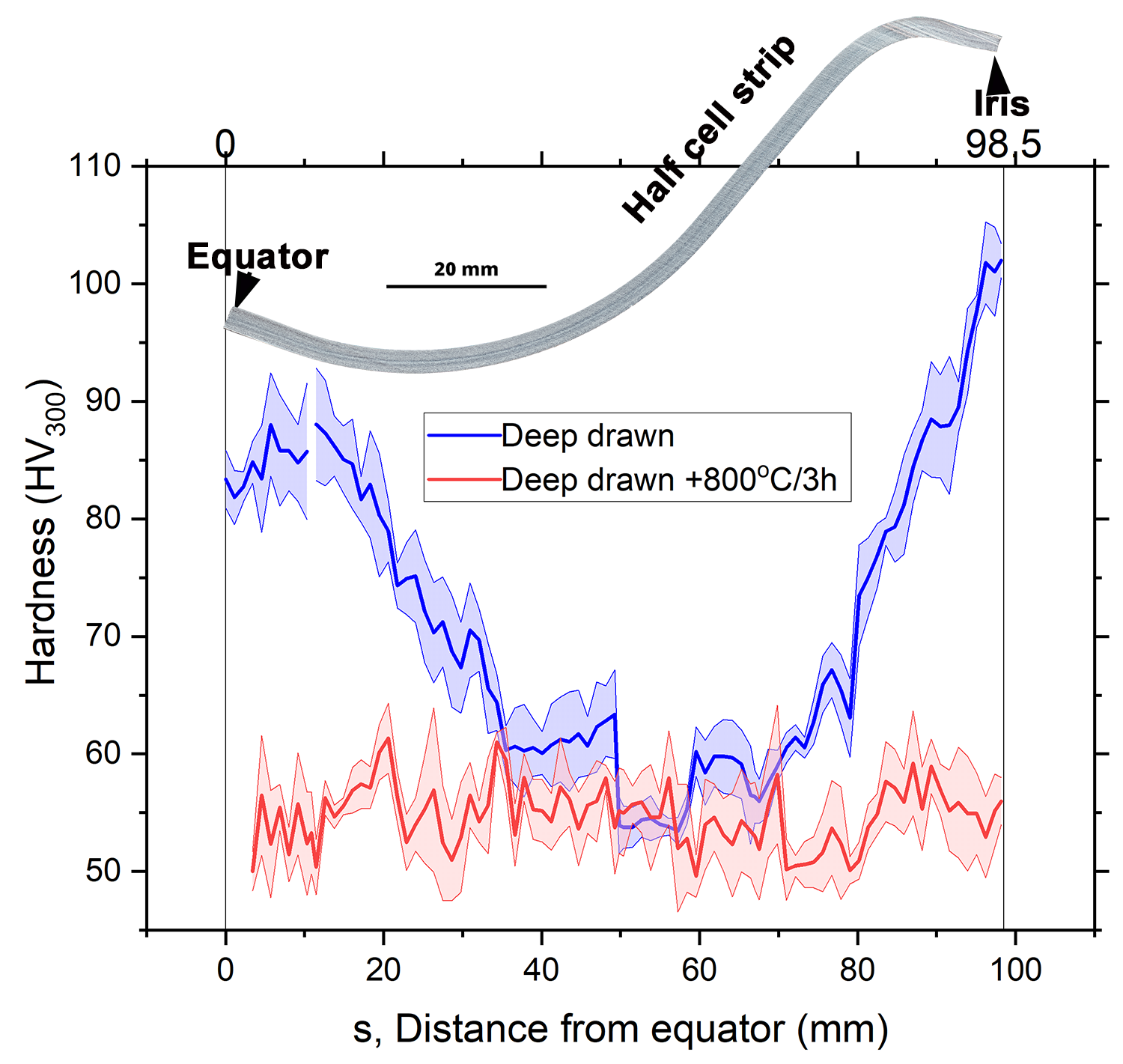}
\caption{\label{fig:hc-trad-hardness} Hardness profile from the equator to iris region from a half-cell fabricated with a traditional SRF Nb sheet with an initial microstructure as shown in Fig. \ref{fig:ebsd} in as deformed state and after 800~$^\circ$C heat treatment. }
\end{figure}

The Nb sheet deformed into a half-cell shape retains characteristics of the deformation process. Figure \ref{fig:hc-trad-hardness} shows the hardness as a function of location along the half-cell cross-section, from the equator to the iris, for the conventional SRF grade Nb. In the deep-drawn state shown by the curve in blue, the highest hardness of HV$_{0.3} \sim$~100 occurs at the iris region, closely followed by the hardness within the first 20~mm of the equator region where the hardness is HV$_{0.3} \sim$~90. There is a gradual drop in hardness from the high hardness regions at the iris and equator towards the center of the half-cell, which is at HV$_{0.3} \sim$ 50. The variation in the hardness as a function of half-cell location after an 800~$^\circ$C/3 h heat treatment shows that the hardness is reduced to a uniformly low level of HV$_{0.3} \sim$~50-60 across the entire length from iris to equator, thus the higher the initial hardness the greater the drop in hardness. At the very center of the half-cell the hardness is the same before and after heat treatment.

\subsection{Development of microstructure after half-cell deformation and 800~$^\circ$C}
Sheets with different initial processing histories (annealed and cold-worked) were drawn into half-cells to simulate the cavity fabrication operation and then annealed at 800~$^\circ$C to reduce the effects of the deformation history and aid in the recovery of the crystalline damage. The analysis focused on the hardness profiles developed for the traditional sheet in Fig. \ref{fig:hc-trad-hardness}. Samples were carefully selected from three different locations: Region a, around the iris region; Region b, 40-50~mm along the profile from the equator region; and Region c, corresponding to 20-30~mm from the equator region. The inverse pole figures (IPF) maps are plotted to show the crystal orientation plane parallel to the half-cell surface in the normal direction (ND). In an SRF cavity, this would be a functional SRF surface in the normal direction. The IPF maps have dark regions (shown in black) indicating a low confidence index (CI $<$ 0.1), implying that the orientations could not be resolved, either due to step size-related resolution of the deformation structure or the sample conditions (i.e surface cleanliness or charging artifacts).

\subsubsection{Traditional Nb sheet formed into half-cell}
\begin{figure}[htb]
\includegraphics*[width=80mm]{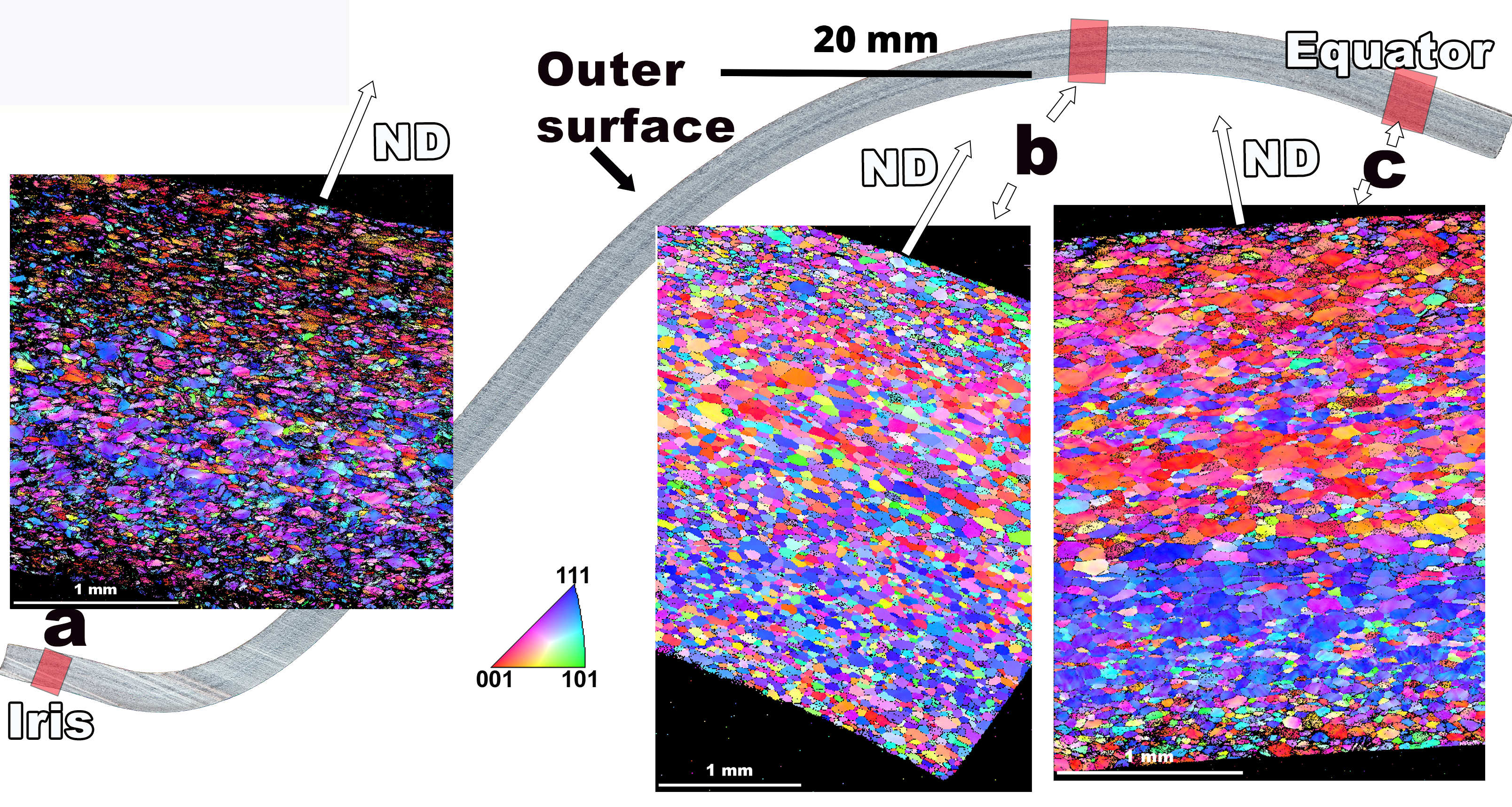}
\caption{\label{fig:trad-worked}IPF maps of selected regions in the ND to the half-cell surface normal showing the variation in deformation in the different areas of the cavity half-cell, (a) close to the iris, which is heavily deformed, (b) and (c) are around the equator regions with lesser deformation. Region c has slightly higher deformation than region b. After half-cell fabrication, the texture gradient is present between the top and bottom surfaces.}
 
\end{figure}

\begin{figure}[htb]
\includegraphics*[width=80mm]{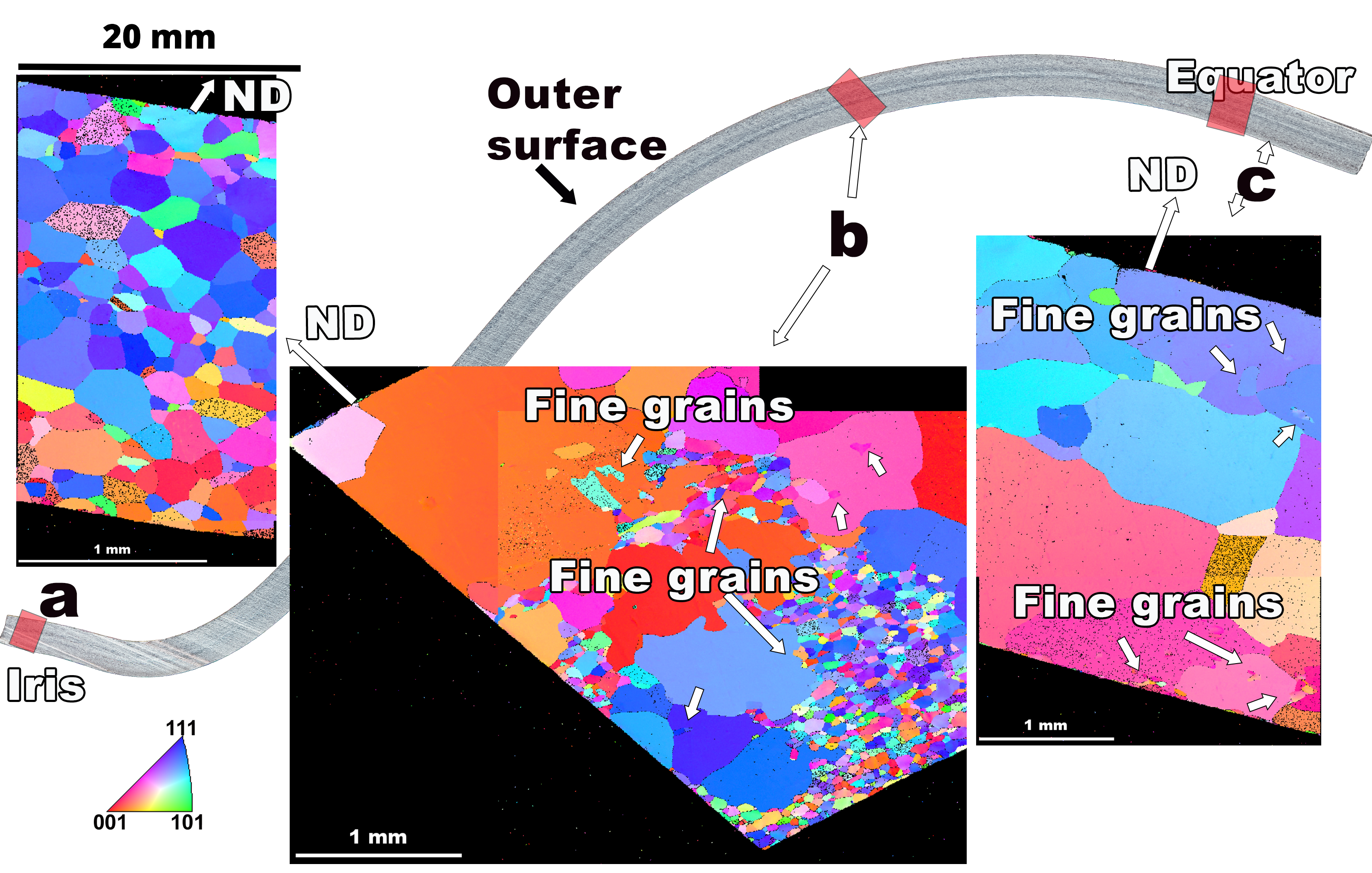}
\caption{\label{fig:trad-worked-800} IPF maps of selected regions in the ND to the half-cell surface normal showing the variations in recrystallization and grain growth in different areas of the cavity half-cell, (a) close to the iris, has fine-grain equiaxed microstructure, and (b) and (c) contain large abnormal grains and finer grains that appear to be trapped inside these large grains.}

\end{figure}

Figure \ref{fig:trad-worked} presents the (IPF) map of a cross-section of a traditional sheet deep-drawn to a half-cell. The significant deformation in the iris region is evident in Fig. \ref{fig:trad-worked} (a), as observed by the in-grain orientation gradients and the change in the aspect ratio of the microstructure (flattened). In contrast, regions b and c exhibit less deformation, where the microstructure of the parent sheet, along with the initial grains, is apparent. Region c, closest to the equator, shows slip traces in the grains, visible in the high-magnification image in Fig.\ref{fig:trad-worked-800}, indicative of light deformation. In all the above IPFs, there are texture gradients across the through-thickness cross-sections.

\begin{figure}[htb]
\includegraphics*[width=80mm]{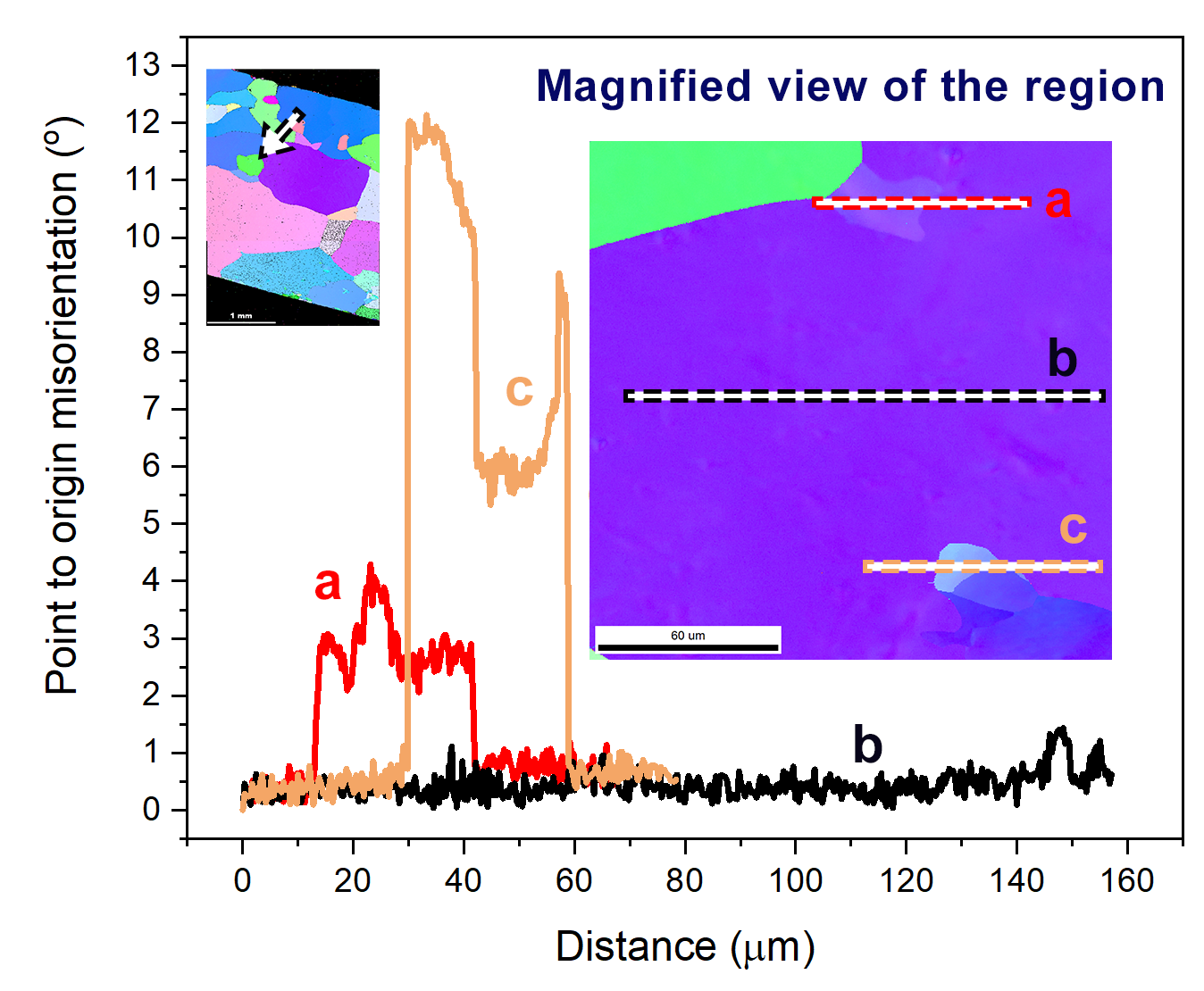}
\caption{\label{fig:highmageq-800} Line scans measuring the misorientation inside the abnormally large grain close to the equator in region c, show the finer details of the abnormal large grains. These regions have local misorientations as low as 2$^\circ$ (a) and as high as 12$^\circ$ (c). The line scan in (b) indicates an expected in-grain line misorientation profile; the variations along the length are the noise floor in the measurement. Note: The IPF map is the same as that of region c in Fig. \ref{fig:trad-worked-800} but plotted in a different color scheme to identify the boundaries visually. }
\end{figure}

 After an 800~$^\circ$C annealing heat treatment, the microstructure changes from the deformed parent structure indicated in Fig. \ref{fig:trad-worked} to a recrystallized and grain growth structure in Fig. \ref{fig:trad-worked-800}. The initially heavily deformed iris region a, has an equiaxed microstructure with an average grain size of 100~$\mu$m. Whereas, the lightly deformed regions, b and c, corresponding to Fig. \ref{fig:trad-worked-800} (b) and (c) show an abnormal grain growth phenomenon with large grains of 100's of $\mu$m and embedded variations in crystal orientations resembling refined grains of tens of micrometers with distinct boundaries. Abnormal grain growth refers to the phenomena of a few large grains that grow abnormally large compared to other surrounding grains in a polycrystalline matrix \cite{humphreys2012recrystallization}. A magnified image of a section of the abnormally grown areas indicates a finer structure, as shown in Fig. \ref{fig:highmageq-800}. There are boundaries inside and along the boundaries of the abnormally grain-grown regions. The point to origin line misorientation in Fig. \ref{fig:highmageq-800} indicates the boundaries encountered are misoriented by $\leq$15$^\circ$ as shown by the line scans in (a) and (c). The line scan in (b) shows the misorientation within the grain crossing no boundaries and can be interpreted as the noise expected with the dataset. 

\subsubsection{Non-traditional sheet formed into half-cell}
\begin{figure}[htb]
\includegraphics*[width=80mm]{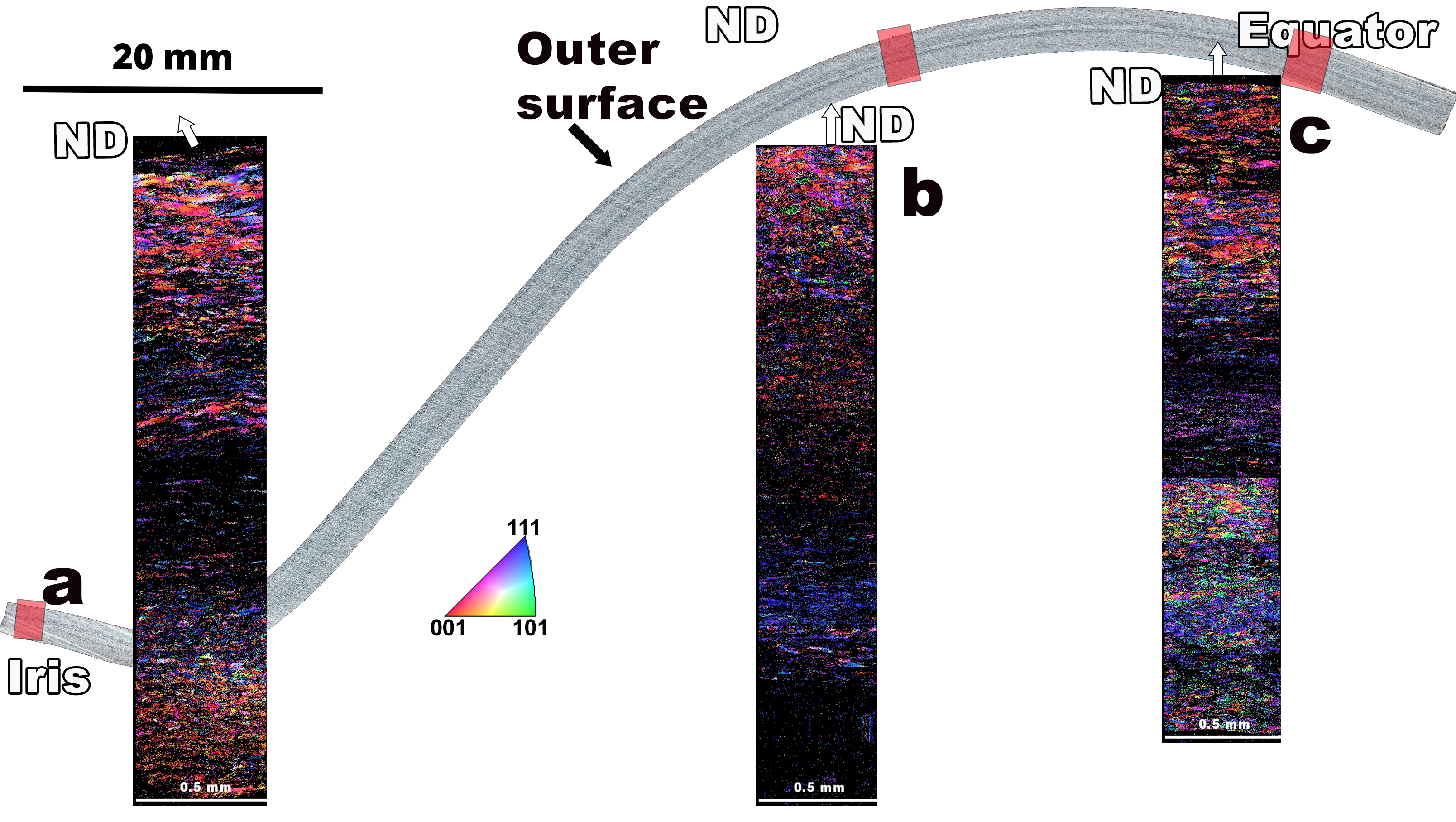}
\caption{\label{fig:ntrad-worked} IPF maps of selected regions in the ND to the half-cell surface normal from the half-cell fabricated with an initially cold-worked sheet showing un-indexed regions in the microstructure corresponding to heavy deformation-related ambiguity determining the crystal orientation with a relatively large step size scans of 1µm. The observed charging artifacts in the signal are the result of the sample being mounted in non-conducting epoxy.}
\end{figure}

\begin{figure}[htb]
\includegraphics*[width=80mm]{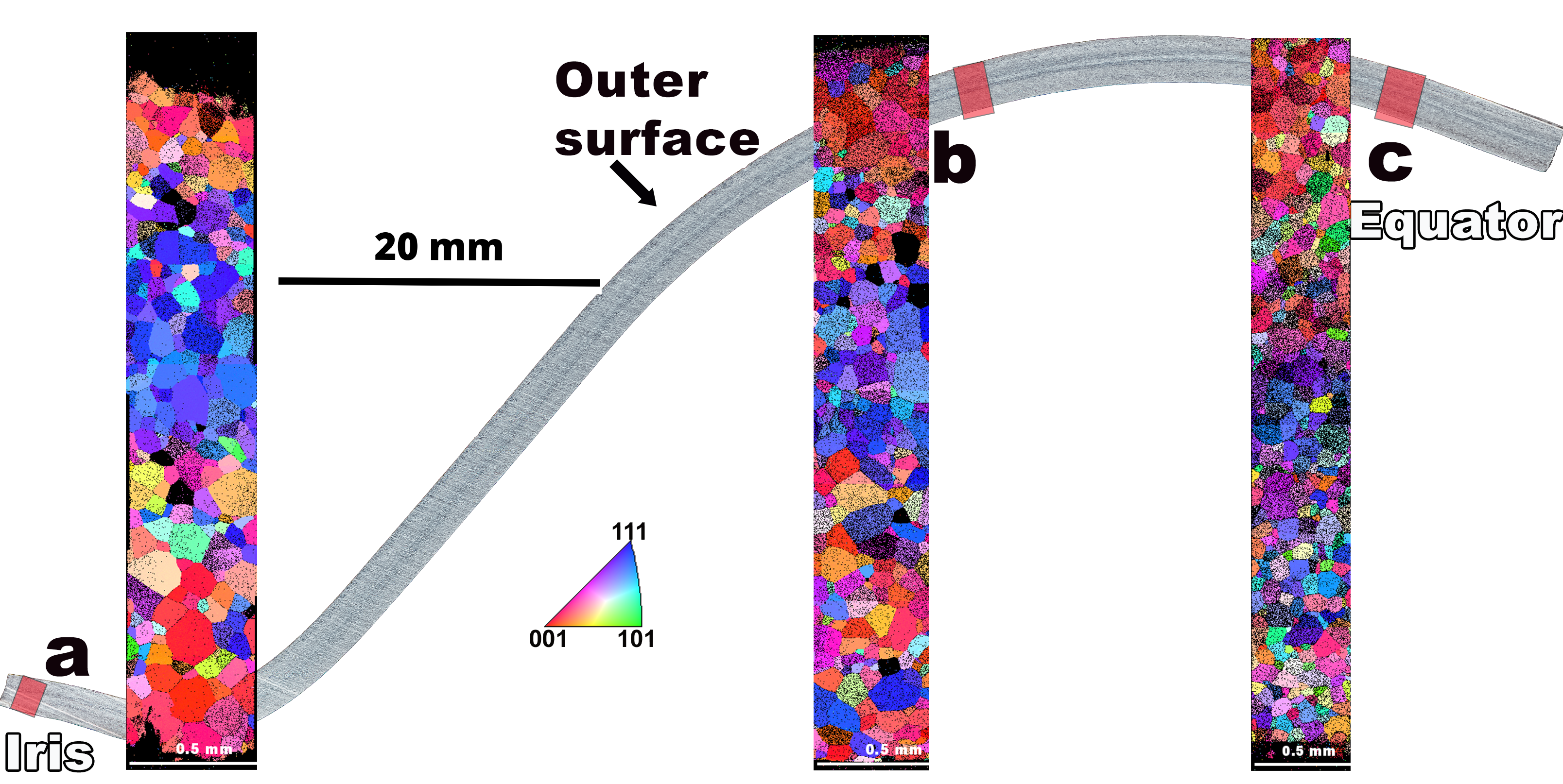}
\caption{\label{fig:ntrad-worked-800} IPF maps of selected regions in the ND to the half-cell surface normal from the half-cell fabricated with an initially cold-worked sheet after 800~$^\circ$C heat treatments showing an equiaxed uniform microstructure in the different regions of the half-cell. The observed charging artifacts in the signal are the result of the sample being mounted in non-conducting epoxy.}
\end{figure}

The half-cell formed from an initial cold-worked sheet must have been better indexed in the different regions analyzed for this study, as shown in Fig. \ref{fig:ntrad-worked}. The detailed deformed microstructure would need higher-resolution imaging of the areas, which is beyond the scope of this study. The purpose is to compare the traditional versus the newer “non-traditional” approach, leading to a higher overall deformation throughout the half-cell. 
After an 800~$^\circ$C heat treatment, the non-traditional-sheet formed half-cell has a more equiaxed microstructure copmared to the traditional-sheet formed half-cell, as seen in Fig. \ref{fig:ntrad-worked-800}. Unfortunately, due to variations in the sample preparation, there were charging issues that were encountered with the non-traditional samples, limiting the direct quantitative comparisons between the traditional and non-traditional sheets in terms of micro-texture and local misorientations. 

\begin{figure}[htb]
\includegraphics*[width=85mm]{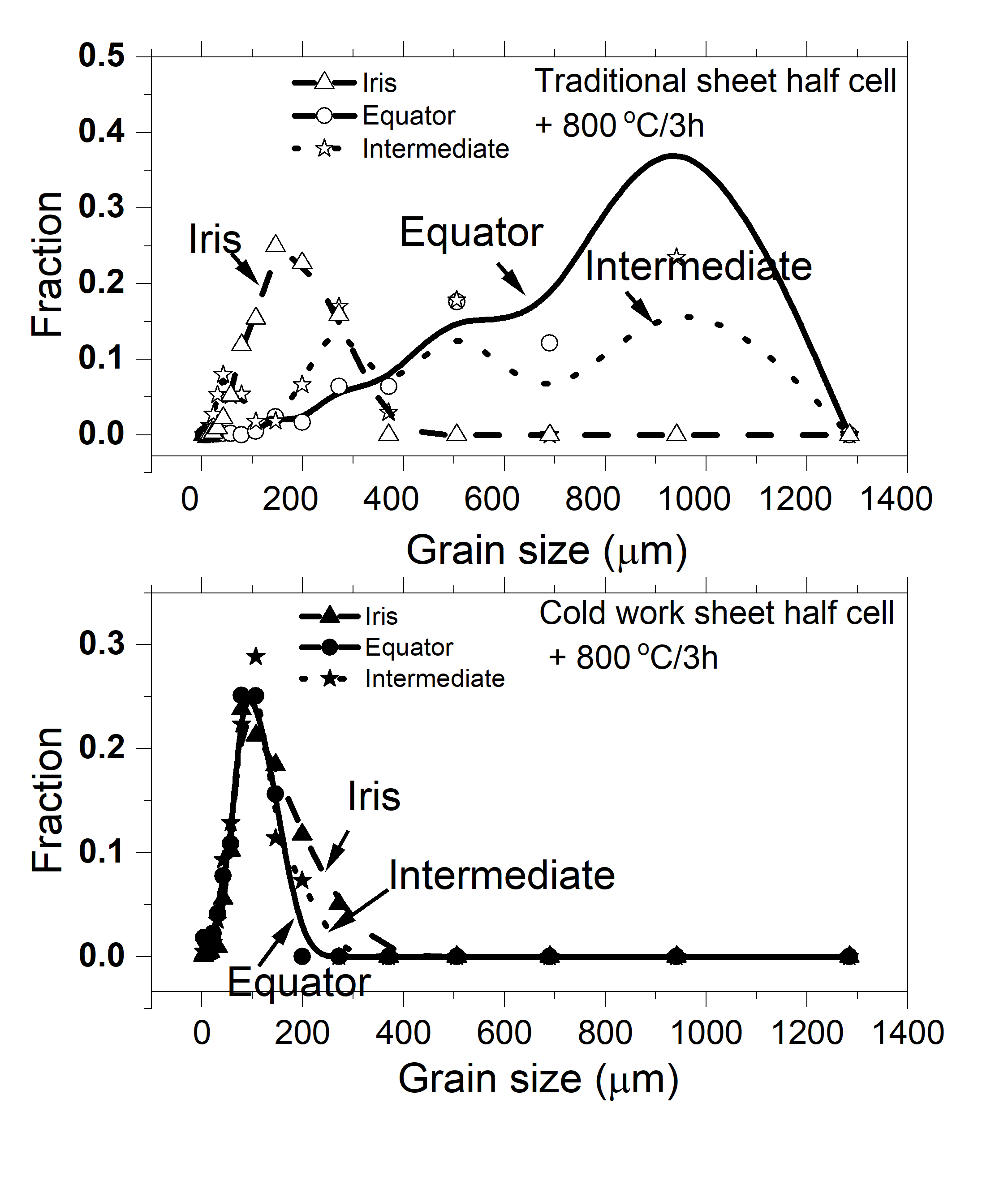}
\caption{\label{fig:gsd}Grain size distributions for the different regions of the traditional sheet formed half-cell indicating multi-modal distributions in the equator and intermediate regions, whereas the non-traditional sheet formed half-cell has similar single-mode distributions for each region, with an average grain size of 100~$\mu$m. }
\end{figure}

However, with the data generated during this study, we have compared the quantitative variations in the grain size after the 800~$^\circ$C heat treatments between the traditional and non-traditional approaches in Fig. \ref{fig:gsd}. The grain size in the iris region a, which was more deformed in the traditional route, results in an average grain size of 100~$\mu$m with a full-width half maximum (FWHM) bandwidth of 75-300~$\mu$m, after heat treatment, whereas the equator and intermediate regions which underwent lower deformation led to abnormal grain growth with a multi-modal grain size distribution centered around 300, 500, and 1000~$\mu$m. In the case of the non-traditional half-cell, we find that the grain size distributions for each region are more uniform and overlap, being centered around 100~$\mu$m. 

\section{\label{sec:sample} Coupon Sample Characterization}
\subsection{Recrystallization of as received cold-worked Nb sheet }
\begin{figure}[htb]
\includegraphics*[width=85mm]{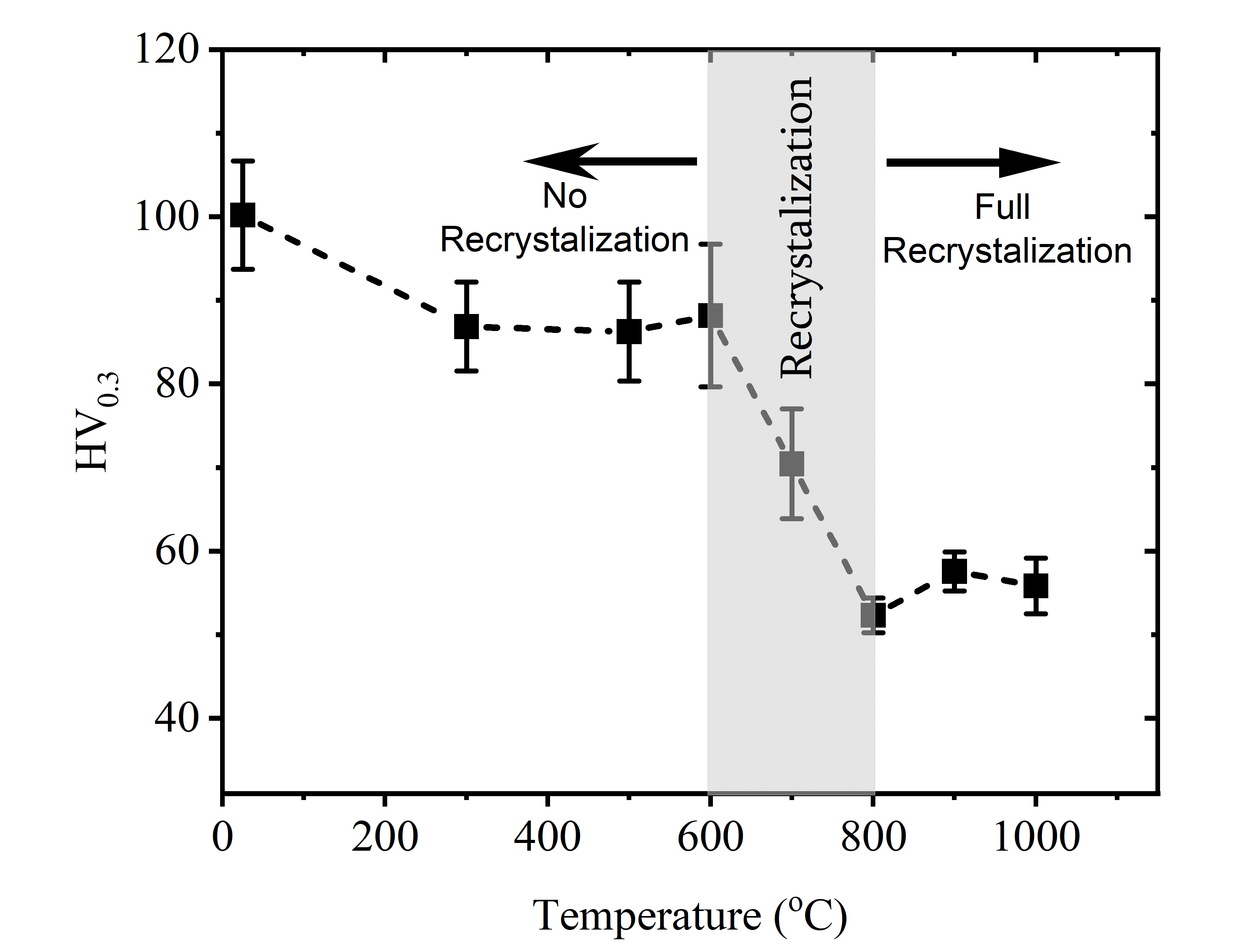}
\caption{\label{fig:HV} Vickers hardness measured on cross-sections of cold-worked Nb as a function of heat treatment temperature. The line is a guide to the eye.}
\end{figure}

The cold-worked Nb sheet from the vendor to form the half-cell was received before the conventional final recrystallization anneal. We observed the recrystallization behavior of the as-received material by plotting the measured hardness (HV$_{0.3}$) as a function of annealing temperature in the range of 300 - 1000~$^\circ$C, at a constant soaking time of 3 hours. The recrystallization curve in Fig. \ref{fig:HV} shows that the hardness decreases as a function of temperature. There is a 10~\% drop in the hardness from the initial cold-work hardness after a 300~$^\circ$C anneal, a 30~\% drop in initial hardness for the 700~$^\circ$C sample, and a nearly 50~\% drop in hardness after the 800~$^\circ$C heat treatment which is maintained after a subsequent 900 - 1000~$^\circ$C heat treatment.

\begin{figure}[htb]
\includegraphics*[width=80mm]{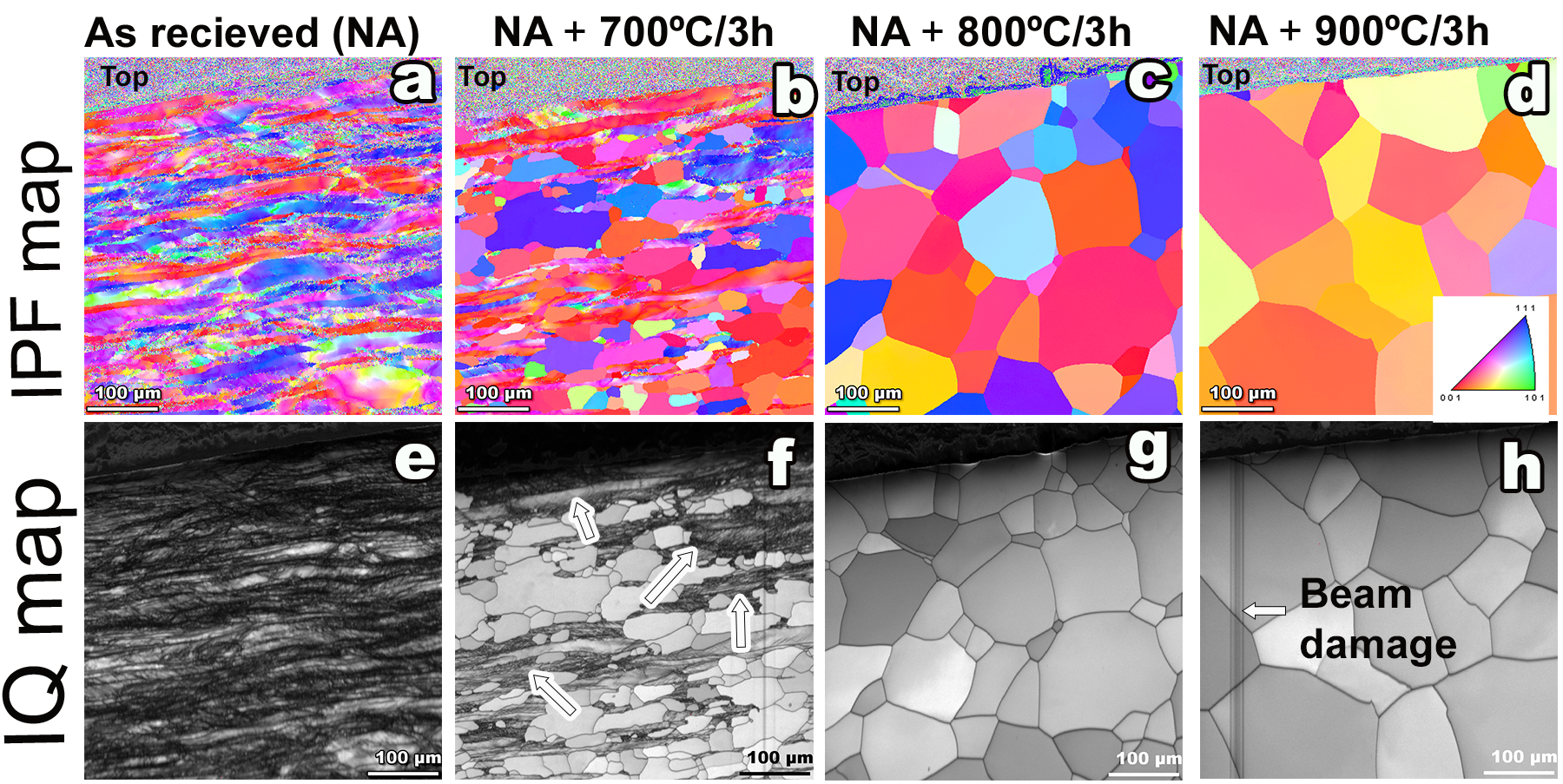}
\caption{\label{fig:texture} Inverse pole figures (IPF), for the sheet cross-sections as a function of heat treatment in the a) non-traditional initial cold-worked sheet, b) after 700~$^\circ$C, c) after 800~$^\circ$C, and d) after 900~$^\circ$C. }
\end{figure}

\begin{figure}[htb]
\includegraphics*[width=85mm]{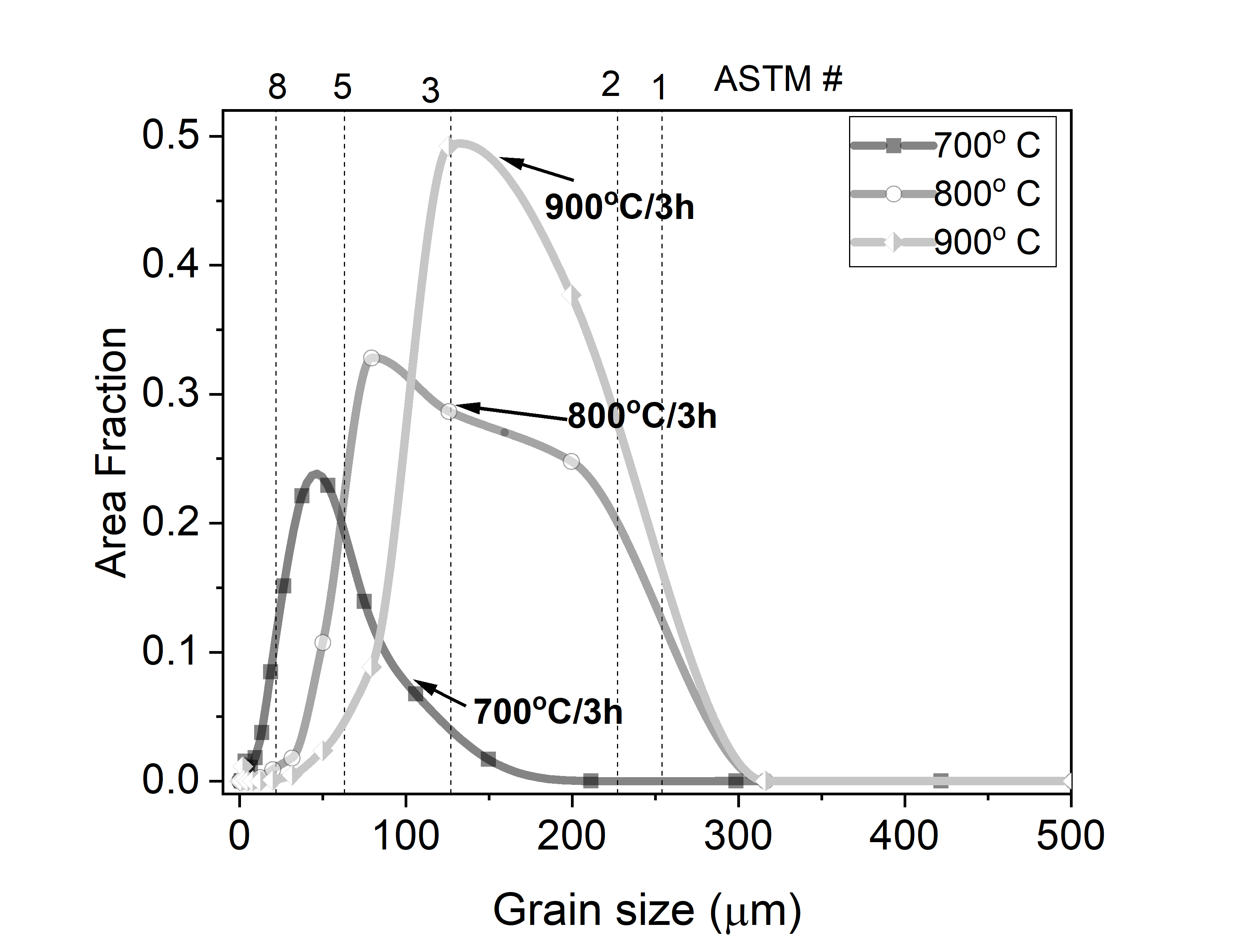}
\caption{\label{fig:gsdsample} Grain size distribution as a function of heat treatment temperature.}
\end{figure}

The corresponding microstructure of the samples after different heat treatment conditions is shown in the IPFs and the image quality (IQ) maps in Fig. \ref{fig:texture}. The cold-worked Nb microstructure in Fig. \ref{fig:texture}. (a) and (e) is one of the elongated grains running parallel to the sheet surface, and the high deformation levels appear dark in the IQ map. After 700~$^\circ$C, new recrystallized grains are visible in Fig. \ref{fig:texture}(b, f) and appear lighter in the IQ map, whereas the deformed microstructure appears darker. After 800~$^\circ$C, there are no deformed regions present in the microstructure images in Fig. \ref{fig:texture} (c,g), and grains are equiaxed with larger grains of the order of 100~$\mu$m, and some finer grains in the range of 20 - 30~$\mu$m. After 900~$^\circ$C, grain growth of the equiaxed grain sizes with more significant fractions of grains in the 100~$\mu$m range are observed in Fig. \ref{fig:texture} (d,f). To quantify the variations in grain size as a function of heat treatment temperature, we plotted the grain size distribution in Fig.\ref{fig:gsdsample}; the average grain size increases from 50~$\mu$m after 700~$^\circ$C to 125~$\mu$m after 900~$^\circ$C, with a bi-modal distribution appearing at 800~$^\circ$C. An interesting observation from this dataset is that the conventional specification for the average grain size at 50~$\mu$m corresponds to a mixture of worked and recrystallized microstructure and not a fully recrystallized one.

\subsection{\label{sec:mag}Thermal Conductivity}
\begin{figure}[H]
\includegraphics*[width=85mm]{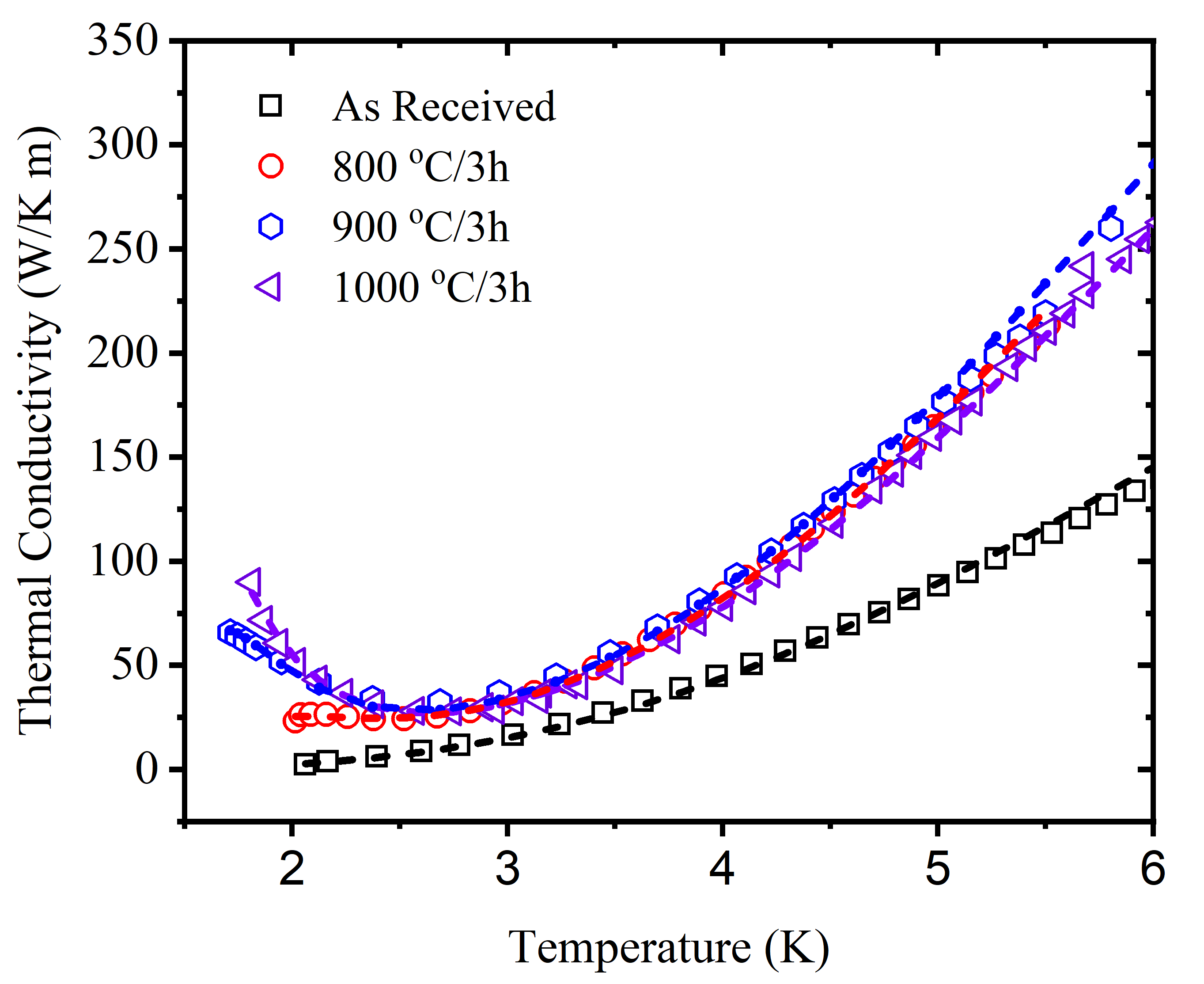}
\caption{\label{fig:thermal} The temperature dependence of thermal conductivity for as-received and heat-treated samples. The dashed lines are fits to thermal conductivity.}
\end{figure}

The temperature dependence of thermal conductivity for different samples under study is shown in Fig. \ref{fig:thermal}. The thermal conductivity of the as-received, cold-worked sample is 2.3~W/K$\cdot$m at 2.0~K and 53~W/K$\cdot$m at 4.2~K, corresponding to a RRR value of 212. As a result of heat treatment, the thermal conductivity at 2.0~K increased to 23.2, 45.3 and 51.6~W/K$\cdot$m after 800~$^\circ$C, 900~$^\circ$C and 1000~$^\circ$C respectively. However, the thermal conductivity at 4.2~K remained around $\sim$ 100~W/K$\cdot$m after the high temperature heat treatment, corresponding to RRR values of $\sim$~400. 
The thermal conductivity data were fitted using the model developed by Xu \cite{xu2019investigation} incorporating randomly distributed dislocation structures. The model predicts the dislocation density by relating the effect of dispersion due to dislocations. Based on the fit the predicted  dislocation density in the as-received material is  $4.7\times 10^{13}$  and decreases to $1.14\times 10^{12}$ after 800~$^\circ$C heat treatment. A previous study on single crystal Nb showed that this change in dislocation density corresponds to the $\sim$ 5~\% uniaxial strain present in as-received sheet\cite{shreyas23}. However, the strain is likely underestimated in cold-worked sheets due to the presence of grain boundaries. Additionally, the phonon mean free path was estimated to be $\sim$~1.75~$\mu$m for as received samples increasing to $\sim$~50~$\mu$m after 800~$^\circ$C heat treatment. The phonon means free path is qualitatively in agreement with the grain size since, at low-temperature, phonons are scattered by grain boundaries. 

\subsection{\label{sec:mag}dc Magnetization and Pinning}
\begin{figure}[htb]
\includegraphics*[width=80mm]{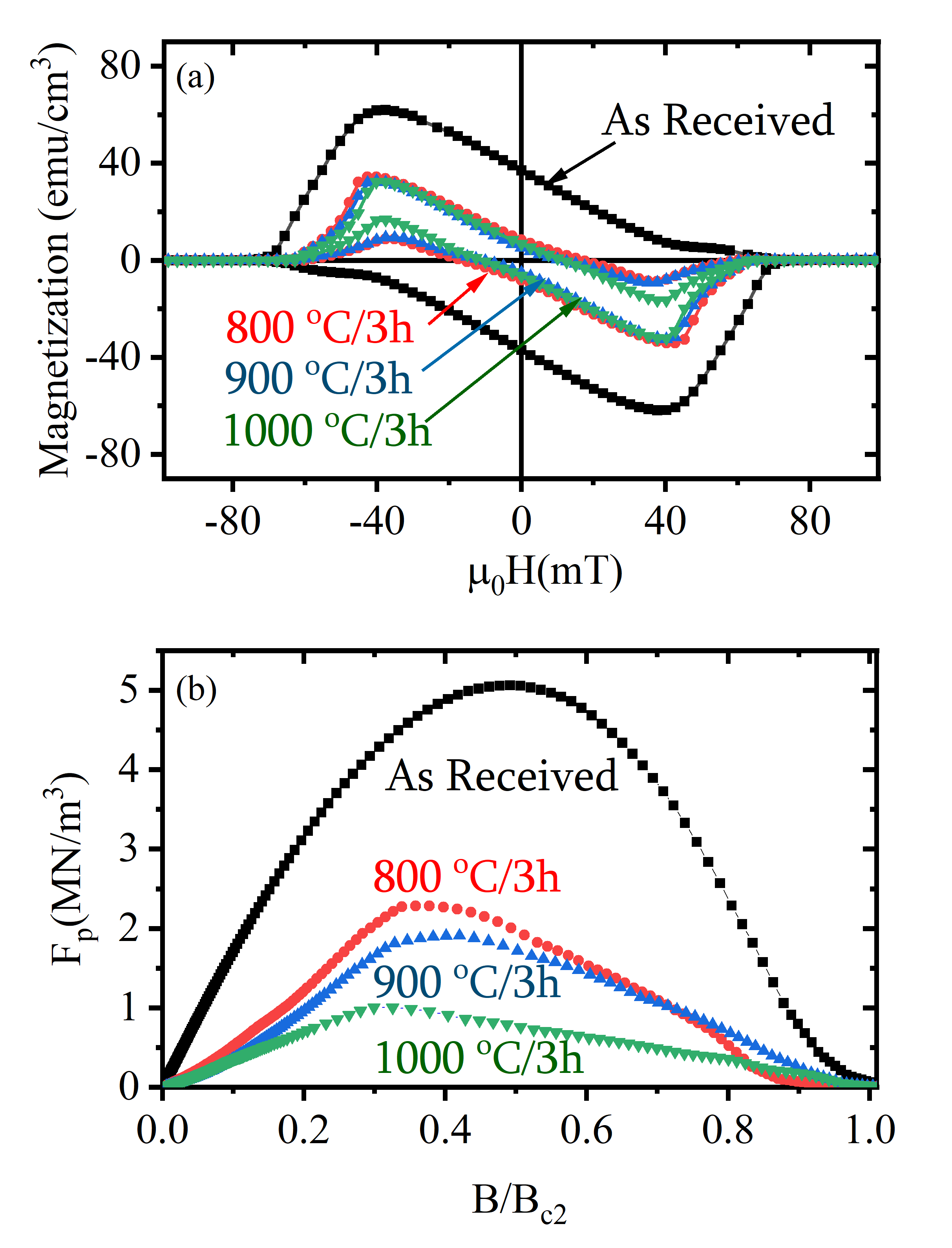}
\caption{\label{fig:pinning} The isothermal dc magnetic hysteresis of the cold-worked Nb samples in zero field cooled state measured at 8.0~K. (b) Pinning force as a function of reduced magnetic field at 8.0~K.}
\end{figure}

The isothermal dc magnetic hysteresis of the samples in zero-field cooled state measured at 8.0~K is shown in Fig.\ref{fig:pinning} (a). The area under the hysteresis loop decreases as we increase the heat treatment temperature, demonstrating the overall reduction in bulk pinning centers from the niobium sample. The pinning force was calculated at 8.0~K and plotted as a function of the reduced magnetic field ($B/B_{c2}$) as shown in Fig. \ref{fig:pinning} (b). The maximum pinning force ($F_p$) of the cold-worked Nb is $\sim$ 5~MN/m$^3$ and it decreases to $\sim$ 2.2~MN/m$^3$ after the 800~$^\circ$C heat treatment. A further decrease in maximum pinning force was observed when the samples were heat treated at higher temperature. The reduction in pinning force is expected due to the removal of pinning sites as a result of heat treatment \cite{dhavale2012flux}. Furthermore, there is no significant change in the pinning force behavior as seen from the location of the peak. The experimental data does not fit with any single pinning model \cite{dew74}, nevertheless the peak at higher reduced field corresponds to the core collective (volume, surface and point) $\delta \kappa$ pinning, where the size of the pinning or the spacing between them are within the penetration depth. The pinning force peak at $B/B_{c2} \sim 0.5$ is mostly dominated by volume pinning, whereas the peak shift towards the lower values $\sim 0.3-0.4$ suggests that the dominating pinning mechanism is point pinning due to the segregation of normal conducting precipitates \cite{dew74}. 

\subsection{\label{sec:mag}Mechanical Properties}
\begin{figure*}[htb]
\includegraphics*[width=170mm]{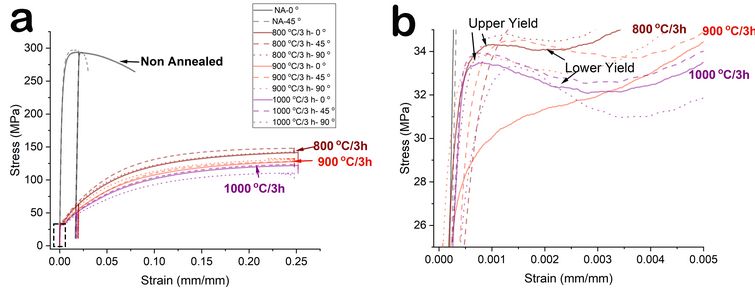}
\caption{\label{fig:mectest} (a) Tensile test curves from the non-traditional cavity sheet after different heat treatments up to 25~\% engineering strain. Notice the significant drop in strength after an 800~$^\circ$C heat treatment. (b) The initial part of the stress-strain curve indicates the phenomena of upper and lower yield point in the annealed material. }
\end{figure*}

\begin{figure}[htb]
\includegraphics*[width=85mm]{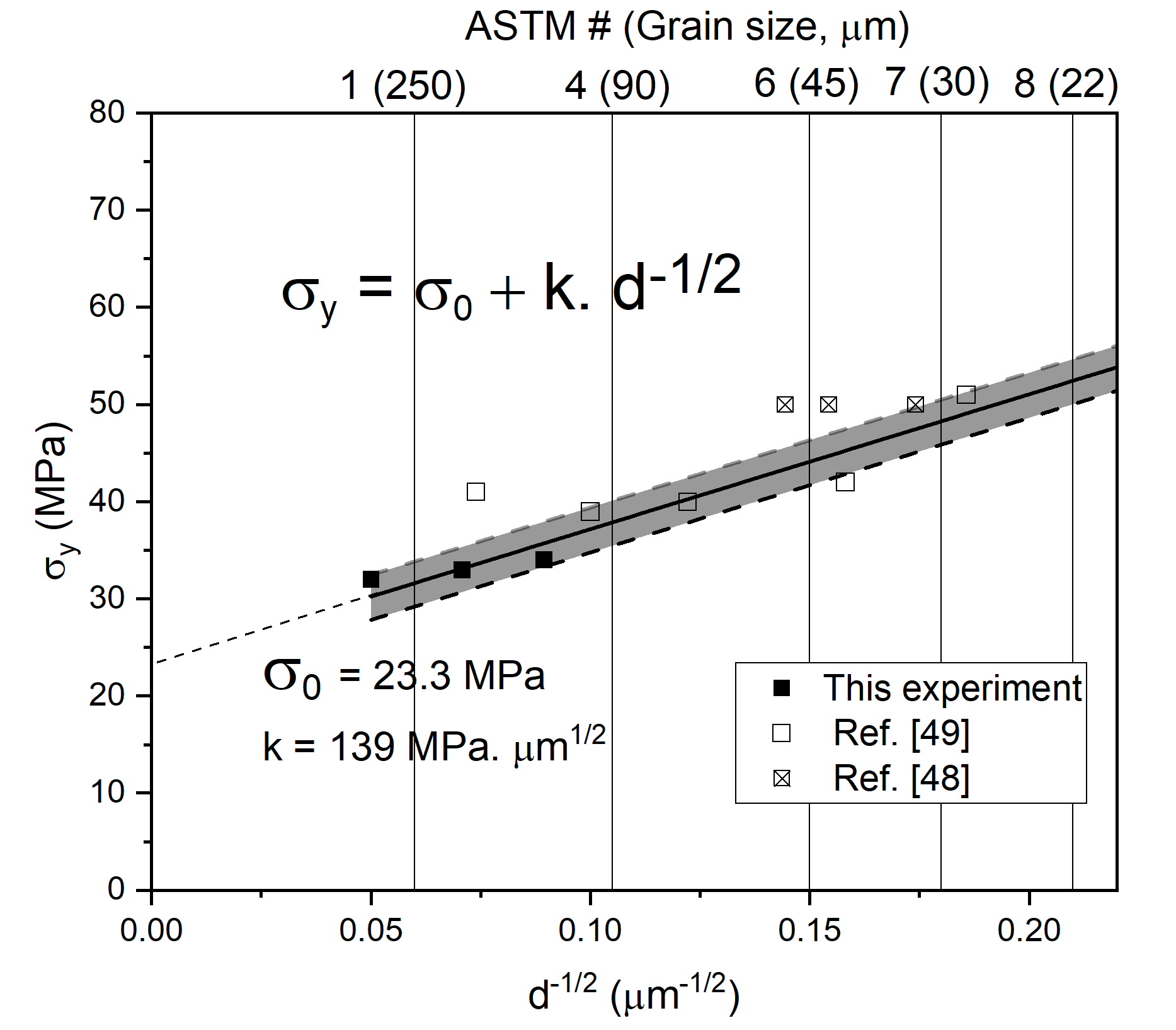}
\caption{\label{fig:hall-petch}Hall-Petch plot of yield strength as a function of grain size is consistent with the expected linear trend and previous studies.}
\end{figure}

Figure \ref{fig:mectest} shows the in-plane tensile tests performed on the non-annealed (NA) samples and after heat treatments to evaluate deformation characteristics in 0$^\circ$, 45$^\circ$, and 90$^\circ$ to the rolling direction (RD). The main difference between the tensile properties of the NA and the heat-treated samples are the following: Material strength is higher in the NA condition versus the heat-treated condition- 800~$^\circ$C. The uniform deformation is lower in the NA state versus the heat-treated samples. The uniform deformation is inferred from the beginning of the drop in stress due to sample necking in the engineering stress-strain curve. In the NA samples, the instability occurs at lower strain values of less than 5~\%. In contrast, instability in the heat-treated samples does not occur beyond a strain of 20~\% as evidenced by the flat-near plastic deformation of polycrystalline Nb.
In the heat-treated samples, the initial deformation has an interesting yield behavior as shown in Fig. \ref{fig:mectest}(b). Most heat-treated tensile test samples have an upper and lower yield point, irrespective of the orientation. To quantify the mechanical properties, stress-strain parameters, which include the elastic modulus (EM), yield strength (YS) at 0.2~\% strain, and the ultimate tensile strength (UTS), and the Vickers micro-hardness (HV$_{0.3}$) is reported in Table \ref{table3}. The modulus of the Nb sheet material varies between an average modulus of 73 - 87~GPa, with the lowest modulus occurring in the NA sheet material. The highest modulus of 89~GPa occurs after a 1000~$^\circ$C heat treatment.
No clear trends are observed between the orientation and in-plane strength values in the directions tested. However, minor variations are observed in the in-plane strength characteristics. The in-plane variation in the YS for the NA sheet is 4~\%, and UTS is 2~\%. In the samples after 800~$^\circ$C, and 900~$^\circ$C the variation in in-plane YS and UTS is less than 3~\%. In the 1000~$^\circ$C sample the variation in the UTS is slightly higher $\sim$7~~\%. 
From the above data, we plot the variation in YS as a function of grain size (d), as shown by the classical Hall-Petch relationship \cite{cordero2016six}, Eq. \ref{hall-petch}.
\begin{equation}
    \sigma_{y}= \sigma_0+ \frac{k}{\sqrt{d}}
\label{hall-petch}
    \end{equation}
where, $\sigma_{0}$ is the material's intrinsic strength related to the critical resolved shear stress, k is the Hall-Petch coefficient, and d is the grain size. 
Figure \ref{fig:hall-petch}, provides the relationship that fits the experimental data with, $\sigma_{0}$= 23.3~MPa, and k = 139~MPa. $\mu$m$^{1/2}$. The obtained YS values are within the 10\% trend lines of the recent measurements on recrystallized SRF cross-rolled Nb sheets \cite{umezawa:ipac2024-thps63, yamanaka2023relation}. Based on the Hall-Petch fit, we can estimate the yield of the material once the grain size is determined. For the cold-worked, as-received material, the average Y.S. is 258~MPa; from the Hall-Petch equation, we obtain an average grain size of 0.35~$\mu$m. However, the Hall-Petch model may not apply to the cold-worked sheets with no well-defined grain sizes.

\begin{table*}[htbp]
  \centering
  \caption{\label{table3}Summary of mechanical properties of the cold-worked sheet}
    \begin{tabular}{cccccccc}
    \toprule
    \multicolumn{1}{p{9.08em}}{{Sample Condition }} & \multicolumn{1}{p{7.34em}}{{Orientation ($^\circ$)}} & \multicolumn{1}{p{6.16em}}{{EM (GPa)}} & \multicolumn{1}{p{5.63em}}{{YS (MPa)}} & \multicolumn{1}{p{5.87em}}{{UTS (MPa)}} & \multicolumn{1}{p{4.13em}}{HV $_{0.3}$} & \multicolumn{1}{l}{YS : HV$_{0.3}$} & \multicolumn{1}{l}UTS : HV$_{0.3}$\\
    \midrule
    \midrule
    Non- annealed (NA) & \multicolumn{1}{c}{0} & 73    & 268   & 294   & \multirow{3}[2]{*}{100 $\pm$ 5} & \multirow{3}[2]{*}{2.6} & \multirow{3}[2]{*}{2.7} \\
          & \multicolumn{1}{c}{45} & 73   & 258    & 298   &       &       &  \\
     & \multicolumn{1}{c}{90} & 73    & 249   & 304   &       &       & \\
    \midrule
    Average  &       & 73      & 258 $\pm$ 10   & 299 $\pm$ 5   &       &       &  \\
    \midrule
    \multicolumn{1}{p{9.08em}}{NA + 800~$^\circ$C} & \multicolumn{1}{c}{0} & 82    & 34    & 144   & \multirow{3}[1]{*}{50 $\pm$ 2} & \multirow{3}[1]{*}{0.7} & \multirow{3}[1]{*}{2.9} \\
          & \multicolumn{1}{c}{45} & 72    & 35    & 149   &       &       &  \\
          & \multicolumn{1}{c}{90} & 82    & 34    & 145   &       &       &  \\
    \midrule
    Average  &       & 79 $\pm$ 6   & 34 $\pm$ 1    & 146 $\pm$ 3  &       &       &  \\
    \midrule
    NA + 900~$^\circ$C & \multicolumn{1}{c}{0} & 84    & 31    & 130   & \multirow{3}[1]{*}{58 $\pm$ 2} & \multirow{3}[1]{*}{0.6} & \multirow{3}[1]{*}{2.2} \\
          & \multicolumn{1}{c}{45} & 75    & 34    & 128   &       &       &  \\
          & \multicolumn{1}{c}{90} & 79    & 33    & 133   &       &       &  \\
    \midrule
    Average  &       & 79 $\pm$ 5   & 33 $\pm$ 2& 130 $\pm$ 3  &       &       &  \\
    \midrule
    NA + 1000~$^\circ$C & \multicolumn{1}{c}{0} & 87    & 32    & 123   & \multirow{3}[1]{*}{56 $\pm$ 1} & \multirow{3}[1]{*}{0.6} & \multirow{3}[1]{*}{2.1} \\
          & \multicolumn{1}{c}{45} & 89    & 31    & 124   &       &       &  \\
          & \multicolumn{1}{c}{90} & 86    & 32    & 110   &       &       &  \\
    \midrule
    Average  &       & 87 $\pm$ 2   & 32 $\pm$ 1   & 119 $\pm$ 8  &       &       &  \\
    \midrule
    \bottomrule
    \end{tabular}%
\end{table*}%
\section{\label{sec:disc}Discussion}

\subsection{Impact of the initial sheet microstructure on cavity deformation and heat treatment}
Figure \ref{fig:hc-trad-hardness} shows the spatial variation in hardness between $\sim$~50 in the mid-section and high hardness of 80-100 in regions close to the equator and iris regions, respectively. As material deforms and yields, the strain-hardening and hardness correlate since the deformation undergone during the deep-drawn cavity profile is not uniform. For SRF Nb, the variation in the hardness provides a proxy for the total strain experienced by the material. The high-strain regions are close to the equator and iris. The hardness is lowered to a fully annealed material level after the 800~$^\circ$C heat treatment. The result of the location-based hardness curve of the annealed sheet is a characteristic of the average macro strain expected in the cavity. 

The results from this study demonstrate that cavities can be manufactured from sheets with retained cold work with no changes to the tooling, dies, or fabrication steps. Initially, cold-worked sheets maintain high levels of prior deformation in the half-cell regions from the equator to the iris and maintain higher dislocation densities that are not indexed with confidence (CI $\geq$ 0.1) within the step size of the limit of micrometers as shown in Fig. \ref{fig:ntrad-worked}.  The cavity formed with a traditional annealed sheet consists of non-uniform cold work as the half-cell shape is formed, as shown by the IPFs in Fig. \ref{fig:trad-worked} between the iris and the equator. The complete cross-section microstructure also indicates that there may be non-uniform deformation within the sheet cross-section in the annealed starting material. 

The strain profile along the cavity half-cell shape directly impacts the microstructural evolution after the 800~$^\circ$C heat treatment. The main difference between the microstructural evolution during heat treatment is:  in the traditional half-cell, a bi-modal microstructure exists throughout the half-cell structure except the heavily deformed iris region. In the moderately and lightly strained regions abnormal grain growth in the order of 100’s of $\mu$m and embedded finer grains in the order of 30 - 50~$\mu$m in the cross-section as shown in Fig. \ref{fig:trad-worked-800} (b), and low angle boundary structures in Fig. \ref{fig:trad-worked-800} (c) occur,  whereas,  equiaxed microstructure in  Fig. \ref{fig:trad-worked-800} (a) occurs in the iris region. The low-angle grain boundaries suggest that the heat treatment temperature cannot fully sweep the stable or pinned low-angle boundaries formed during the recovery stages. This leads to an incomplete recrystallized grain \cite{sakharov2022analysis}. The abnormal grain growth at 800~$^\circ$C in very high purity Nb in regions of low strain may be related to a lack of nucleation points due to low strains and impurities leading to a few grains, possibly due to favorable orientations growing rapidly over others. In high-strain regions, a more uniform distribution of nucleation sites for new grains could lead to a more uniform grain size.

In the half-cell fabricated with cold-worked sheets, the level of deformation is high prior to the deep drawing. After heat treatment, a more uniform microstructure is observed as shown in Fig. \ref{fig:ntrad-worked-800}, with the grain size curves from different regions overlapping around a mean value of $\sim$100~$\mu$m as shown in Fig. \ref{fig:gsd}. The tails of the curves in Fig. \ref{fig:gsd} are slightly different among the other regions, indicating that even with the available initial cold work in this study combined with the half-cell deformation, the grain growth characteristics may still be somewhat different between the regions. The iris region c, has an overall sharper distribution compared to the region a, and region b. In terms of the strain path (hardness) for the  half-cell deformation the strain in region a $>$ region c $>$ region b. These results prove that a critical amount of cold work is necessary to homogenize the grain size in cavity cross-sectional microstructures after 800~$^\circ$C heat treatment. Otherwise, the cavity deformation, being spatially non-uniform, leads to a difference in microstructure, leading to regions of bi-modal large-fine grain structures within the sheet cross-section. The amount of critical cold work that is needed is unknown, but the vendor provided these sheets before any heat treatment step at the upstream processing step. 

\subsection{Impact of microstructure on flux expulsion.}
Figure \ref{fig:expulsion} showed that the cavity fabricated from cold-worked Nb sheet has improved flux expulsion characteristics after 800~$^\circ$C compared to a cavity made from traditional Nb sheet. The variation in flux expulsion behavior has been previously correlated to average grain size \cite{sung2023evaluation}, implying that the larger the grain size the higher the flux expulsion. Here,  we show that the average grain size in the traditional sheet after 800~$^\circ$C heat treatment is location dependent and non-uniform, i.e. finer grains embedded in large grains as shown in Figs.\ref{fig:trad-worked-800} and \ref{fig:highmageq-800}. The average grain size is $\sim$ 150~$\mu$m at iris and as high as 1~mm at the equator as shown in Fig. \ref{fig:gsd}. In the case of the cold-worked sheet cavity after 800~$^\circ$C, the average grain size is $\sim$ 100~$\mu$m. However, after the 800~$^\circ$C the flux expulsion performance is superior in the cavity fabricated with cold-worked sheet versus the traditional sheet. It is possible to achieve the full flux expulsion limit on cavities fabricated with cold work Nb after 800~$^\circ$C heat treatment if one can maintain the temperature gradient between the cavity irises above 0.25~K/cm, which is not the case with cavities made from conventional SRF grade Nb. As shown in Table \ref{table2}, the flux trapping sensitivity does not change between the different starting sheets that have undergone similar surface treatments, consistent with the literature \cite{dhakal20flux}. Fundamentally, our results show that the flux expulsion directly correlates to the non-uniformity of grain size rather than the average grain size. Previously, we have observed that a non-uniform, bi-modal grain size distribution leads to flux trapping and poor flux expulsion in fine grain regions \cite{balachandran21}. These cavity results and the cavity cut-out microstructure strengthen the hypothesis that bi-modal, non-uniform fine-large grain structures possibly caused by inadequate recrystallization tend to show poorer flux expulsion.
 
The shape of the flux expulsion curve is different depending on the temperature, irrespective of the initial sheet condition. For the 800~$^\circ$C case the flux expulsion ratio increases approximately linerly with the temperature gradient, whereas for the 900~$^\circ$C and 1000~$^\circ$C has a different behavior, showing a rapid expulsion of flux as the temperature gradient is increased. We interpret these results as related to the underlying microstructure and subtle variations between the surface and bulk grains. The interior surface of Nb cavity is highly susceptible to damage, and this damaged layer could influence the flux expulsion performance \cite{claireprab}. Based on recent results by Thune and co-workers \cite{thune2023influence, thuneinfluence, balachandran2023microstructure}, surface regions within the first few microns of the SRF niobium may recrystallize differently from the bulk, as suggested by the decreased surface recrystallized fractions after 800~$^\circ$C and 900~$^\circ$C heat treatments. Whereas, the bulk recrystallizes 100~\% after 800~$^\circ$C, which is also confirmed from the coupon study experiments of the hardness-based recrystallization curve in Fig. \ref{fig:HV}. The grain size distribution in Fig. \ref{fig:gsd} suggests a finer grain peak that is below 100~$\mu$m. The OIM in Fig. \ref{fig:texture}(c), suggests that there are fine grains in the regions approximately within 10~$\mu$m from the surface. Other researchers have also observed surface abnormality in grain growth \cite{martinello2019microscopic, sung2023evaluation}, and this is the first interpretation directly linking to flux expulsion. We do not understand the exact nature of why these fine grains are present but speculate that dislocations may be blocked by surface barriers such as oxides \cite{sun2023surface}, carbides\cite{dangwal2021grain, sun2023surface, chen2024unraveling}, or other contaminants from the heat treatment furnace. It is well known that 800~$^\circ$C dissolves the Nb pentoxide effectively \cite{semione2019niobium}. The variations between the 800~$^\circ$C and higher temperatures may be related to the inability of the dislocations that are getting annealed to go past the oxide barrier at low temperatures, leading to a surface zone of non-recrystallized or fine grain size layer. Gaining more statistics from the surface layers and comparing them with bulk microstructure as a function of temperature and chemistry can unequivocally answer this question.
We can also infer from this result that decreasing the grain boundary density with heat treatment reduces the overall number of pinning sites, leading to a change in the shape of the flux expulsion curve. The pinning force curves in Fig. \ref{fig:pinning} show that as the heat treatment temperature increases, the pinning force drops, and there is a slight shift in the pinning force curve to the left. An open question remains as to whether variations in the dc pinning force curves indicate subtle changes in the pinning mechanism or are they dominated by surface effects.

\subsection{Summary of coupon studies and relevance to SRF Nb cavities}
Coupon studies provide a test bed to evaluate fundamental microstructural-physical property correlations relevant to SRF cavities. The coupon studies presented here provide quantitative microstructure data that could correlate with the cavity performance which is affected by flux expulsion or pinning, thermal conductivity, and early flux penetration. The mechanical property data could also provide engineering design guidelines that may help with pressure vessel code designs. 
In this paper, following the recrystallization steps of an initial vendor-supplied non-annealed sheet, we find grain size dependence on recrystallization, mechanical strength, thermal conductivity (hence RRR), and pinning force curves. From our results here, Figs. \ref{fig:texture} and \ref{fig:gsd}, specification of finer grain size of 50~$\mu$m or lesser leads to an initial material that is not completely recrystallized. The impact of this incomplete initial recrystallization after deformation leads to low-strained regions in the cavities having dislocation structures solely due to the initial sheet condition rather than cavity deformation. As shown in Fig. \ref{fig:highmageq-800}, these remnant dislocation structures or lightly deformed regions can lead to bi-modal microstructure due to abnormal grain growth and retain dislocation structures that could decrease the Nb's ability to expel magnetic flux.

The thermal conductivity measurements indicate clear differences between non-annealed and recrystallized Nb, especially at 2~K, where the phonon modes are critical for heat transfer. However, at 4.2~K, there are no significant changes in the heat transfer characteristics. The RRR as inferred from the non-annealed sheet after 1000~$^{\circ}$C varies between 200 to 400 depending on the initial sheet condition. Assuming that the sheet chemistry (mainly determined by the number of electron beam melting steps) does not change, we find that the variation in thermal conductivity and hence the RRR is a function of microstructure. The results here indicate that with increases in boundary density either due to a cold-worked state or low heat treatment temperature, the RRR can vary significantly. This can be expected in the range of 200- 400 in a cavity after deformation and heat treatment. The dc-magnetization and pinning force curves in Fig. \ref{fig:pinning} also suggest the variation that is possible in the strength of the pinning force locally. 
The mechanical properties in Table \ref{table3}, and Figs. \ref{fig:mectest} and \ref{fig:HV} indicate the variation in strength as a function of grain size and heat treatment, which needs consideration, especially in designing for pressure vessel codes \cite{peterson2010pure, ciovati2015mechanical}. From the Hall- Petch equation for Nb determined here, large surface grains of the order of millimeters like those in Fig \ref{fig:trad-worked-800} lead to local soft spots due to a decrease in YS. This decrease in local yield strength could lead to cavity shape changes during handling, directly impacting cavity tuning. These large/bi-modal grains have also been observed before during the SNS project (2000s) and LCLS-II (present) \cite{myneni2003elasto, sung2023evaluation}
In general, keeping the impurity content constant, the strength is a function of grain size for a given purity of SRF Nb. An essential consideration for SRF is surface roughness, and quantifying the average surface roughness in the uniform deformation region of the various samples shows the influence of the initial microstructure on the roughness. Different areas of cavities may need different amounts of removal to achieve the same smoothness. Implications of this varying roughness based on grain size that may exist apriori in Nb sheets may lead to regions of different roughness when optimization based on dirty layer removal is prioritized \cite{PhysRevAccelBeams.22.122002}. The open question is the impact of surface roughness on the enhancement of flux penetration, and recent work suggests that topographic defects substantially suppress the superheating field and, hence, in the reduction of an achievable accelerating gradient in SRF cavities \cite{Ericprab_24}. From the half-cell deformation studies, the non-annealed sheet develops a uniform microstructure in different regions of the cavity, and the impact of the varying physical, electromagnetic, and thermal properties is likely to be minimized. 

\subsection{Adoption of cold-worked sheet into mainstream cavity fabrication}
Cold-worked sheets benefit the flux expulsion behavior, particularly at low heat treatment temperatures, by aiding better recrystallization. The strength of SRF Nb depends primarily on the grain size or grain boundary density. The lower the grain size, the higher the GB density and the better the strength as indicated in Fig. \ref{fig:gsd}, and Table \ref{table3}. Choosing lower heat treatment temperatures and obtaining superior performance, i.e. high $Q_0$, by lowering the residual resistance is possible in SRF cavities from any vendor if a cold-worked sheet is chosen over a traditional sheet. 
However, adopting this strategy for cavity manufacture will require exploring the formability ranges of cold-worked Nb sheets. A uniaxial test shows a low uniform deformation of 5\% at 45$^\circ$ in-plane to the rolling direction and 10\% along the rolling direction, suggesting low in-plane deformability. Nb is highly formable and can be simulated by parametrized models to predict bi-axial deformation. The extent of deformability depends on the strain rate, and servo controls available \cite{kim2022mechanical}. The lack of uniform deformation could be an issue with cavity shapes and needs exploration of the formability diagram \cite{hecker1975simple}. Annealed sheets show uniform work hardening and good formability beyond 25\% in-plane after heat treatments of the initial cold-worked sheet.

\section{\label{sec:concl}Summary}
We have successfully fabricated and tested 1.3 GHz TESLA-shaped cavities made with non-traditional sheets with retained cold-work, and lower uniform in-plane deformation. Half-cell experiments, using the standard deformation path for such cavities, reveal continuous hardness variations from the equator to the iris, corresponding to the variation in the deformation path during deep drawing. These deformation path variations lead to microstructural variations after 800~$^\circ$C heat treatment. Specifically, there exist bi-modal grain size distributions, with regions containing fine (50-100~$\mu$m) and large (greater than 100~$\mu$m) grains. The flux expulsion and trapping sensitivity variability was measured on cavities made from cold worked (non-annealed) and traditional SRF grade Nb. We find better flux expulsion after heat treatment in cavities made from cold-worked Nb compared to the cavity made from traditional annealed SRF grade Nb \cite{khanalsrf23}. The direct correlation between the importance of uniform recrystallization and flux expulsion is established. No variations were observed in terms of $Q_0(E_{acc})$ and flux trapping sensitivity as long as the final surface preparation remains the same. Adopting a new sheet strategy would need further exploration including the development of new specifications, as well as additional optimization focused on maximizing the microstructural benefits of using non-annealed sheet. Several single cell cavities were fabricated using different level of cold-work Nb sheet and results will be made available in future publications. 

\begin{acknowledgments}

We acknowledge Jefferson Lab technical staff members for the cavity fabrication, processing, cryogenic, and rf support. The work done at Florida State University is supported by the U.S. Department of Energy, Office of Science, Office of High Energy Physics under Awards No. DE-SC 0009960 (FSU) and the State of Florida. Additional support for the National High Magnetic Field Laboratory facilities is from the NSF: NSF-DMR-2128556. Work done by M. Barron at Jefferson lab was supported by the U.S. National Science Foundation Research Experience for Undergraduates at Old Dominion University Grant No. 1950141. This is authored by Jefferson Science Associates, LLC under U.S. DOE Contract No. DE-AC05-06OR23177.

\end{acknowledgments}

\bibliography{coldworkNb}% Produces the bibliography via BibTeX.

\end{document}